\documentclass[%
 reprint,
superscriptaddress,
 amsmath,amssymb,
 aps,
prb,
]{revtex4-2}

\usepackage{xcolor}
\usepackage{graphicx}% Include figure files
\usepackage{dcolumn}% Align table columns on decimal point
\usepackage{bm}% bold math
\usepackage[colorlinks=true,citecolor=blue,linkcolor=magenta]{hyperref}
\usepackage{cases}

\def \mc{\mathcal}

\begin{document}

\preprint{APS/123-QED}

\title{Generalized Rashba electron-phonon coupling and superconductivity in strontium titanate}

\author{Maria N. Gastiasoro}
\email{maria.ngastiasoro@dipc.org}
\affiliation{ISC-CNR and Department of Physics, Sapienza University of Rome, Piazzale Aldo Moro 2, 00185, Rome, Italy}
\affiliation{Donostia International Physics Center, 20018 Donostia-San Sebastian, Spain}
\author{Maria Eleonora Temperini}%
\affiliation{ISC-CNR Institute for Complex Systems, via dei Taurini 19, 00185 Rome, Italy}%
\author{Paolo Barone}
\affiliation{SPIN-CNR Institute for Superconducting and other Innovative Materials and Devices, Area della Ricerca di Tor Vergata, Via del Fosso del Cavaliere 100, 00133 Rome, Italy}
\author{Jos\'e Lorenzana}
\email{jose.lorenzana@cnr.it}
\affiliation{ISC-CNR and Department of Physics, Sapienza University of Rome, Piazzale Aldo Moro 2, 00185, Rome, Italy}%

\date{\today}% It is always \today, today,
             
\begin{abstract}
SrTiO$_3$ is known for its proximity to a ferroelectric phase and for showing an “optimal” doping for superconductivity with a characteristic dome-like behaviour resembling systems close to a quantum critical point. Several mechanisms have been proposed to link these phenomena, but the abundance of undetermined parameters prevents a definite assessment. Here, we use {\em ab initio} computations supplemented with a microscopic model to study the linear coupling between conduction electrons and the ferroelectric soft transverse modes allowed in the presence of spin-orbit coupling.  
We find a robust Rashba-like coupling, which can become surprisingly strong for particular forms of the polar eigenvector. We characterize this sensitivity for general eigenvectors and, for the particular form deduced by hyper-Raman scattering experiments, we find a BCS pairing coupling constant of the right order of magnitude to support superconductivity. The {\em ab initio} computations enable us to go beyond the linear-in-momentum conventional Rashba-like interaction and naturally explain the dome behaviour including a characteristic asymmetry. The dome is attributed to a momentum dependent quenching of the angular momentum due to a competition between spin-orbit and hopping energies. The optimum density for having maximum $T_c$  results in rather good agreement with experiments without free parameters.  
These results make the generalized Rashba dynamic coupling to the ferroelectric soft mode a compelling pairing mechanism to understand
bulk superconductivity in doped SrTiO$_3$. 
\end{abstract}

\maketitle
%\tableofcontents

\section{Introduction}

A research surge in recent years has uncovered novel behavior involving the interplay between ferroelectricity (FE) and superconductivity (SC) in SrTiO$_3$ (STO)~\cite{Edge2015,collignon2019}.
Noteworthy examples include strain enhanced superconductivity~\cite{Herrera2019,Ahadi2019} in samples with polar nanodomains~\cite{salmani2021role,salmani2021interplay} and self-organized dislocations with enhanced ferroelectric fluctuations~\cite{hameed2020ferroelectric}. 
Alternative methods for tuning ferroelectricity such as Ca or $^{18}$O isotope substitution also present enhanced superconducting critical temperatures $T_c$~\cite{Stucky2016,Rischau2017,Tomioka2019,tomioka2022,rischau2022isotope}. 
In doped samples with a global polar transition (and thus global broken inversion symmetry) signatures of mixed-parity superconductivity have been reported~\cite{schumann2020possible}. Theoretically, these doped polar samples have also been recently proposed as a platform for the emergence of exotic phases such as Majorana-Weyl superconductivity~\cite{yerzhakov2022majorana} and odd-frequency pair correlations~\cite{kanasugi2020ferroelectricity}.

Despite having experimentally established a qualitative connection between the superconducting and ferroelectric phases in STO, and while there is some indication that the dominant mode responsible for pairing might be the ferroelectric soft transverse optical (TO) mode~\cite{franklin2021giant}, there is still no consensus about the pairing mechanism in this system~\cite{Gastiasoro2020review}. One the of the prominent theoretical challenges is its very low density of states and Fermi energy due to low carrier densities, which places superconductivity in STO outside of the standard BCS paradigm. 

Proposed pairing theories include the dynamical screening of the Coulomb interaction due to
longitudinal modes~\cite{Takada1980,Ruhman2016,wolfle2019reply,Enderlein2020} recently challenged in Ref.~\cite{edelman2021normal}, bipolaron formation~\cite{lin2021analysis}, and diverse approaches to linear coupling~\cite{Edge2015,Arce-Gamboa2018quantum,kedem2018novel,kozii2019,Gastiasoro2020,Sumita2020,yoon2021low,gastiasoro2022theory,yu2021theory,kozii2021synergetic} or quadratic coupling to the FE mode~\cite{Ngai,van2019possible,kiseliov2021theory,volkov2021superconductivity,zyuzin2022anisotropic}. The last two proposals have the advantage that, coupling electrons directly to the FE soft mode, provide a natural explanation to the sensitivity to the FE instability.

In the more general context of polar or nearly polar metals, the coupling between electrons and the soft FE modes has received attention only very recently~\cite{kozii2019,Volkov2020,Gastiasoro2020,kozii2021synergetic,kumar2022spin,gastiasoro2022theory,zyuzin2022anisotropic,klein2022theory}. The reason probably being that, as already mentioned, the FE soft modes in these systems have a predominantly \emph{transverse} polarization. Within the conventional electron-phonon interaction scheme, this implies a decoupling of the soft modes from the electronic density to linear order~\cite{ruhman2019comment}. 
One promising alternative route involves going to next order by coupling the electrons to pairs of TO modes, {\em i.e.} the quadratic coupling mentioned above~\cite{Ngai,van2019possible,kiseliov2021theory,volkov2021superconductivity,kumar2021}. Another possibility, and subject of the present article, is the linear \emph{vector} coupling to the electrons, allowed in the presence of spin-orbit coupling (SOC)~\cite{Fu2015,Kozii2015,Martin2017,kozii2019,Kanasugi2019,kanasugi2020ferroelectricity,Gastiasoro2020,kumar2022spin}.

In a recent work~\cite{gastiasoro2022theory} we derived a vector coupling based on a Rashba-like interaction within a minimal microscopic model and \emph{ab initio} frozen phonon computations. The interaction originates from a combination of inter-orbital coupling to the inversion breaking polarization of the mode and SOC~\cite{Petersen2000}.  
In the minimal model we assumed a conventional Rashba coupling linear in the electronic momentum $k$. 
In the present work we show that this approximation is valid only at very low densities. Because of this, the problem of the dome in STO could not be addressed in Ref.~\cite{gastiasoro2022theory}.

Here, we present a complete study of the spin-orbit assisted coupling between the low-energy electronic bands and the FE soft TO modes in tetragonal doped STO. We find that the magnitude of the Rashba coupling is strongly sensitive to the particular form of the eigenvector of the soft mode. Indeed, we discover a gigantic coupling to the polar mode deforming the oxygen cage, 
so that even a small admixture of this distortion in the eigenvector of the soft mode makes the coupling to electrons
quite large. Furthermore, we find that a naive, linear-in-$k$ Rashba coupling deviates strongly from the {\em ab initio} computations when the electronic wave-vector exceeds a small fraction of the inverse lattice constant. Incorporating these results into a generalized Rashba coupling and using the soft-mode eigenvector deduced from hyper-Raman scattering,  we find a dome-like behavior of the superconducting $T_c$ with a maximum value of the correct order of magnitude. The origin of the dome can be explained with a minimal model of generalized Rashba coupling. Also the position of the dome maximum and its characteristic asymmetry as a function of doping are in good agreement with experiment without free parameters. 
Our work shows that a generalized Rashba pairing mechanism explains bulk SC in doped STO. 
We refer here to the standard definition of bulk superconductivity as the one which shows the Meissner effect. 

This mechanism may also be relevant in two-dimensional electron gases at oxide interfaces~\cite{liu2021two,chen2021electric,mallik2022superfluid}. It has recently been proposed that the extreme sensitivity of superconductivity to the crystallographic orientation of KTaO$_3$ (KTO) can be explained by invoking the linear coupling to TO modes~\cite{liu2022tunable}. KTO is also an incipient ferroelectric, and hence the coupling to the soft FE mode may be important for pairing as well.

The paper is organized as follows. In Sec.~\ref{sec:electronic} we introduce the multiband electronic structure of STO, which is successfully described by a tight-binding model fit to {\em ab initio} band-structure computations within Density Functional Theory (DFT). Because we are interested in coupling the electrons to zone-center polar phonon modes, in Sec.~\ref{sec:Si} we  present a complete basis $\bar S_i$ to parametrize any polar mode belonging to tetragonal $E_u$ and $A_{2u}$ irreducible representations (irreps).
In Sec.~\ref{sec:linear} we show how a linear-in-$k$ Rashba-like coupling between the electrons and zone-center polar modes emerges from a microscopic model in the presence of SOC, and estimate the coupling constants with the aid of \emph{ab initio} frozen-phonon computations in STO. 
The corresponding electron-polar-phonon coupling Hamiltonian is then derived in Sec.~\ref{sec:el-ph}; we find all three electronic bands have a substantial dynamic Rashba coupling to the soft TO mode in STO. In Sec.~\ref{sec:SC} we use the {\em ab initio} results and a minimal model to explore the superconducting properties derived from the generalized Rashba mechanism. 
%, and for the phase diagram as a function of carrier density. 
%The estimated pairing constants $\lambda_\mathrm{BCS}$ can provide the right order of magnitude of $T_c$, and develop a dome when increasing the Fermi wave vector due to deviation from Rashba linear-in-$k$ coupling. 
We finally present our conclusions in Sec.~\ref{sec:conclusions}.

\section{Electronic structure}
\label{sec:electronic}
\subsection{Electronic DFT bands}
%The low-energy electronic band structure of STO computed by DFT (see \cite{SM} for details), shown in Fig.~\ref{fig:bands} (dashed lines), consists of three  doubly degenerate bands around the zone center.
%Below 105 K, an antiferrodistortive (AFD) structural transition lowers the symmetry of STO from cubic [space group $Pm\bar{3}m$, Fig.~\ref{fig:bands} (a)] to tetragonal  [space group $I4/mcm$, Fig.~\ref{fig:bands} (b)], which results in a split of the lower two bands at the zone center. At even lower temperatures, upon electron doping STO, superconductivity develops at a few hundred $m$K. 
We first discuss the electronic band structure of STO as computed by DFT.  We adopted the projector augmented-wave (PAW) method as implemented in VASP~\cite{vasp1,vasp2} and the Perdew-Burke-Ernzerhof generalized gradient approximation revised for solids (PBEsol)~\cite{pbesol}.
An antiferrodistortive (AFD) structural transition is known to occur below 105 K, therefore we considered both the high-temperature cubic (space group $Pm\bar{3}m$) and the low-temperature tetragonal (space group  $I4/mcm$) unit cell. We first relaxed both structures until forces were smaller than 1 meV/\AA, using a plane-wave cutoff of 520 eV and a Monkhorst-Pack grid of 8$\times$8$\times$8 and 6$\times$6$\times$6 k-points for cubic and tetragonal phases, respectively. Optimized lattice constants are $a=3.907$ \AA~ for cubic STO and $a_t=5.508$ \AA, $c_t=7.845$ \AA~ for tetragonal STO. Electronic structure calculations have then been performed with the inclusion of SOC, as implemented in VASP\cite{steiner2016}.

\begin{figure}
    \centering
    \includegraphics[width=\linewidth]{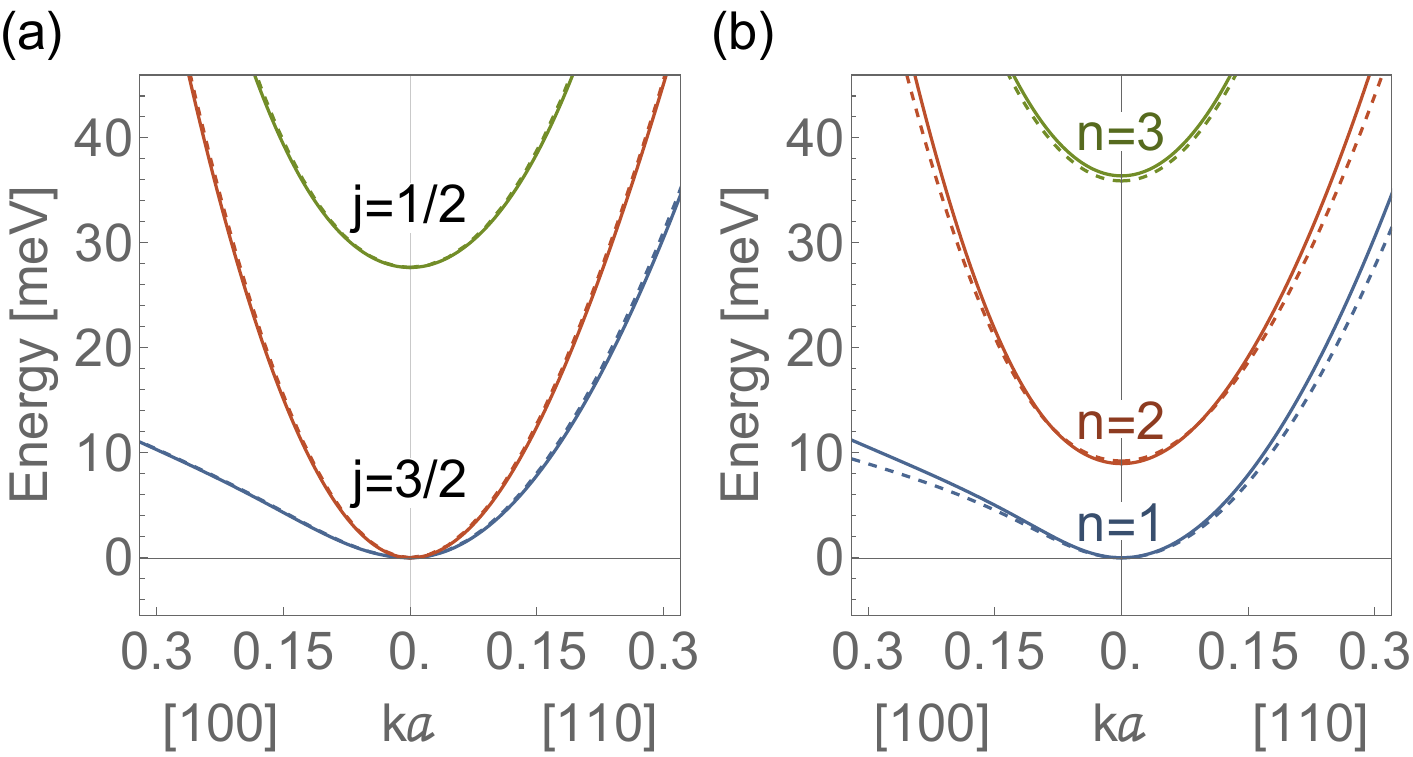}
    \caption{Low energy band structure of STO in the (a) cubic state and (b) tetragonal state. Dashed lines are computed from DFT and full lines the tight-binding model in Eq.~\eqref{eq:tb-model}. The momenta in (a) and (b) are along the cubic and pseudocubic directions respectively, and $a$ is the cubic lattice constant. }
    \label{fig:bands}
\end{figure}

The low-energy electronic band structure is shown in Fig.~\ref{fig:bands} (dashed lines) and consists of three  doubly degenerate bands around the zone center. AFD distortions result in a split of the lower two bands at the zone center, as displayed in Fig.~\ref{fig:bands} (b). 

Superconductivity develops upon electron doping the tetragonal STO at a few hundred $m$K. The resulting superconducting state spans the filling of the three bands shown in  Fig.~\ref{fig:bands} (b) before vanishing~\cite{collignon2019}, starting from a zero-resistance state in the very dilute single-band regime with a Fermi energy of a few meV, and evolving into bulk multi-band SC with a Fermi energy of a few tens of meV.  

\subsection{Minimal electronic model}\label{sec:minmod}
A minimal tight-binding model with the $3d$ $t_{2g}$ orbitals of the Ti atom $yz$, $zx$ and $xy$, denoted respectively $\mu=x,y$, and $z$ in this work, successfully describes the low-energy electronic band dispersion (full lines in Fig.~\ref{fig:bands}) in both cubic and tetragonal phases~\cite{Bistritzer2011,vanderMarel2011,Zhong2013,Gastiasoro2020review,gastiasoro2022theory}. 
%Here we adopt the fit of Appendix D of Ref.~\cite{gastiasoro2022theory}. 
The non-interacting model Hamiltonian reads, 
\begin{equation}
    \mc{H}=\mc{H}_0+\mc{H}_\mathrm{SOC}+\mc H_\mathrm{AFD}.
    \label{eq:tb-model}
\end{equation}
Here, we have included a hopping term up to next-nearest neighbors,
\begin{equation}
    \mc{H}_0=\sum_{\bm{k} s \mu\nu}t_{\mu\nu}(k)c^\dagger_{\mu s}(\bm{k}) c_{\nu s}(\bm{k})
\end{equation} 
between orbitals $\mu$ and $\nu$ with spin $s=\pm1/2$ (shorthanded as $s=\pm$ in operator labels as in $c^\dagger_{\mu +}$).  
The atomic SOC of the $t_{2g}$ manifold reads, 
%$\mc{H}_{SOC}=\xi\sum_{\mu s\nu s' l}i \epsilon_{\mu\nu l}\sigma_{l,ss'}c^\dagger_{\mu,s}c_{\nu,s'}$
\begin{equation}\label{eq:so}
 \mc{H}_\mathrm{SOC}=-2\xi \bm{l}\cdot \bm{s}   
\end{equation}
where we introduced an \emph{effective} orbital moment operator with $l=1$~\cite{sugano1970,khomskii2020orbital,stamokostas2018mixing}. The physical orbital angular momentum $\bm{l}(t_{2g})$ is directed in the opposite direction with respect to ${\bm l}$. 

Finally,  the tetragonal crystal field term is, 
\begin{equation}\label{eq:adf}
 \mc H_\mathrm{AFD}=\Delta\sum_{\mu s} \delta_{\mu,z}c^\dagger_{\mu s} c_{\mu s}   
\end{equation}
 which effectively accounts for the AFD distortion by shifting the energy of the $\mu=z$ orbital~\cite{gastiasoro2022theory}.
 
The $t_{2g}$ hopping term reproduces the low-energy quadratic dispersion given by DFT along the high-symmetry cubic directions [Fig.~\ref{fig:bands}(a)], 
\begin{align}
     t_{\mu\mu}(\bm{k})=&-2t_1\left(\cos k_\alpha+ \cos k_\beta \right)-2 t_2 \cos k_\mu \nonumber\\
     &-4 t_3 \cos k_\alpha \cos k_\beta+(4t_1+2t_2+4t_3)\label{eq:tmumu}\\
   t_{\mu\nu}(\bm{k})=&-4 t_4 \sin k_\mu \sin k_\nu,\label{eq:tmunu}
 \end{align}
with hopping parameters $t_1=451$ meV, $t_2=40$ meV, $t_3=111$ meV and $t_4=27$ meV. In Eq.~\eqref{eq:tmumu} $\alpha\neq\beta$ and $\alpha,\beta\neq \mu$, while in Eq.~\eqref{eq:tmunu},  $\mu \neq \nu$.

The eigenstates of Eq.~\eqref{eq:so} can be classified with  an effective total angular momentum $\bm j=\bm l+\bm s$ with the associated quantum number $j$.
The SOC term breaks the six-fold degeneracy of the $t_{2g}$ manifold at the zone center, opening a $3\xi=28$ meV gap between the lower multiplet $j=3/2$ and the higher doublet $j=1/2$ in the high-$T$ cubic state [Fig.~\ref{fig:bands}(a)].
In terms of the $t_{2g}$ orbital operators $c^\dagger_{\mu s}$ these new eigenstates and associated operators, $c^\dagger_{j,j_z}$,  take the following form at the zone center~\cite{khomskii2020orbital}:
\begin{align}
\label{eq:c3232}
    c^\dagger_{\frac{3}{2},\pm\frac{3}{2}}&=\mp\frac{1}{\sqrt{2}}\left(c^\dagger_{x,\pm}\pm i c^\dagger_{y,\pm}\right)\\
    c^\dagger_{\frac{3}{2},\pm\frac{1}{2}}&=\frac{1}{\sqrt{6}}\left(\mp c^\dagger_{x,\mp}- i c^\dagger_{y,\mp}+ 2 c^\dagger_{z,\pm}\right)\label{eq:c3212}\\
    c^\dagger_{\frac{1}{2},\pm\frac{1}{2}}&=\frac{1}{\sqrt{3}}\left(-c^\dagger_{x,\mp}\mp i c^\dagger_{y,\mp}\mp  c^\dagger_{z,\pm}\right).
    \label{eq:c1212}
    %\psi^\dagger_{\frac{3}{2},\frac{3}{2}}&=\frac{1}{\sqrt{2}}\left(\psi^\dagger_x+i\sigma_3 \psi^\dagger_y\right)\\
    %\psi^\dagger_{\frac{3}{2},\frac{1}{2}}&=\frac{1}{\sqrt{6}}\left(\sigma_x\psi^\dagger_x-\sigma_y\psi^\dagger_y-2\sigma_z\psi^\dagger_z\right)\\
    %\psi^\dagger_{\frac{1}{2},\frac{1}{2}}&=\frac{1}{\sqrt{3}}\left(\sigma_x\psi^\dagger_x-\sigma_y\psi^\dagger_y+\sigma_z\psi^\dagger_z\right)
\end{align}
The tetragonal crystal field term $\mc H_\mathrm{AFD}$ in Eq.~\eqref{eq:tb-model}, does not affect $j_z=\pm 3/2$ states which remain therefore eigenstates of the full Hamiltonian at the zone center. Instead, it mixes states with $j_z=\pm 1/2$ (i.e. states with non-zero $\mu=z$ orbital character) and thus splits the degeneracy of the lowest multiplet $j=3/2$ at the zone center. A fitting to the DFT band structure in the low-$T$ tetragonal phase [Fig.~\ref{fig:bands}(b)] gives $\Delta=17.7$ meV, and sets the following order of the three doubly-degenerate bands at $\Gamma$:  
 \begin{align}
 \label{eq:tetbasis1}
 c^\dagger_{1,\pm} &=c^\dagger_{\frac{3}{2},\mp\frac{3}{2}}\\\nonumber
 \mc E_{1}&=0 \\
 c^\dagger_{2,\pm} &=\cos \theta c^\dagger_{\frac{3}{2},\pm\frac{1}{2}} \mp\sin \theta c^\dagger_{\frac{1}{2},\pm\frac{1}{2}}\\\nonumber
 \mc E_{2}&=\frac{1}{2}\left(3\xi+\Delta-\sqrt{9\xi^2-2\xi\Delta+\Delta^2}\right) \\
  \label{eq:tetbasis3}
 c^\dagger_{3,\pm} &=\pm\sin\theta c^\dagger_{\frac{3}{2},\pm\frac{1}{2}} + \cos \theta c^\dagger_{\frac{1}{2},\pm\frac{1}{2} }\\\nonumber
 \mc E_{3}&=\frac{1}{2}\left(3\xi+\Delta+\sqrt{9\xi^2-2\xi\Delta+\Delta^2}\right)  
 \end{align}
 from lowest to highest energy $\mc E_1<\mc E_2<\mc E_3$, and with $\tan 2\theta=\frac{-2\sqrt{2}\Delta}{-\Delta +9\xi}$.
 %See Ref.~\cite{SM} for details. 
Note that the pseudospin index $\pm$ of the bands in Eqs.~\eqref{eq:tetbasis1}-\eqref{eq:tetbasis3} is chosen to coincide with the projection of the electronic spin along the real orbital moment instead of the effective orbital moment within the T-P equivalence~\cite{stamokostas2018mixing} ($\bm{l}(t_{2g})=-\bm{l}$, see also~\cite{SM}).

Carrying the analysis for general momentum we can write the electronic Hamiltonian Eq.~\eqref{eq:tb-model} in the absence of a polar distortion as
\begin{equation}
 \mc{H}= \sum_{n  \bm{k} }  \psi_n^\dagger(\bm{k}) \mc E_{n}(\bm{k}) \sigma_0 \psi_{n}(\bm{k})   
 \label{eq:Hband}
\end{equation}
where we defined the spinor $\psi^\dagger_n=(c^{\dagger}_{n+},c^{\dagger}_{n-})$ for band $n$, and introduced the $2\times2$ identity matrix $\sigma_0$ for pseudospin degeneracy.  Figure~\ref{fig:bands}(b) shows that this model with the parameters quoted above gives an excellent fit of the bands obtained by DFT in the presence of both AFD and SOC.

\section{Polar soft mode in STO}
\label{sec:Si}

\begin{figure*}
    \centering
    \includegraphics[width=\linewidth]{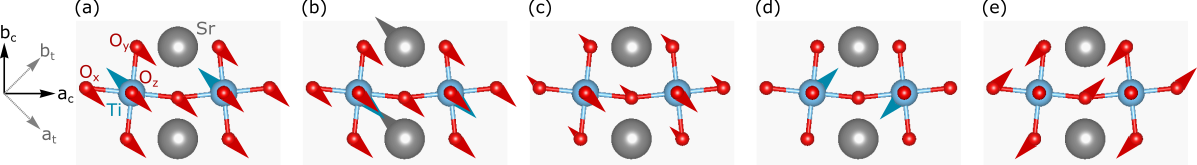}
    \caption{Atomic distortions of the $E_u$ irrep for $Ima2$ ($C_{2v}$) with polar axis $\bm{\hat{n}}_p \parallel [1\bar{1}0]$ for basis modes (a) $\bar S_1$, (b) $\bar S_2$, (c) $\bar S_3$, (d) $\bar S_4$ and (e) $\bar S_5$ given by Eqs.~\eqref{eq:S1}-\eqref{eq:S5} and Table~\ref{tab:Si-modes}. In all cases distortions are on the plane of the drawing and the arrowheads identify the atoms moving. The left inset shows the pseudocubic and tetragonal in-plane axes. The Wickoff positions of Ti, O$_x$ and O$_y$ atoms are given in~\cite{SM}. 
    The atomic displacements for the basis of $A_{2u}$ mode with polar axis $\bm{\hat{n}}_p\parallel [001]$ are equivalent to panels (a)-(c) with all displacements pointing out of the plane, i.e. rotated by $\pi/2$ around the $b_t$ tetragonal axis. }
    \label{fig:Si-sketch}
\end{figure*}

Although we are focusing on the temperature region where an AFD is present, it is customary to discuss the atomic displacements of the near zone-center polar soft mode in terms of a complete set of basis modes defining symmetry coordinates for the $T_{1u}$ irrep of $O_h$ in the high-$T$ cubic phase.
Indeed, Axe~\cite{1967Axe} introduced one such possible set of coordinates to describe the eigenvectors of polar normal modes in cubic perovskite structures, which has been used to restrict the possible atomic distortions of the various polar modes that were compatible with reflectivity~\cite{1967Axe}, neutron scattering~\cite{Harada1970} and hyper-Raman experiments~\cite{Vogt1988}. Since this coordinate set has been widely used and referred to in the literature of polar modes in STO, we shall use it in our work as well.  
Within this framework, a general polar distortion can be decomposed into symmetry coordinates in the following way
\begin{align}
\bm{ \bar U}=( \bm{r}^\mathrm{Sr},\bm{r}^\mathrm{Ti},\bm{r}^{\mathrm{O}_x},\bm{r}^{\mathrm{O}_y},\bm{r}^{\mathrm{O}_z})=\sum_i \bm{\hat{n}}_i u_i \bar S_i.\label{eq:U}\end{align}
Here, $\bm{\hat{n}}_i$ is a unit vector setting the direction of atomic displacements for basis mode $i$. We shall see that, in general, displacements with the same polar axis, $\bm{\hat{n}}_p$, do not need to be collinear; this requires a different  $\bm{\hat{n}}_i$ for each basis mode. $ \bar S_i$ defines the basis of eigenmodes expressed in terms of collinear atomic displacements $(s^\mathrm{Sr},s^\mathrm{Ti},s^{\mathrm{O}_x},s^{\mathrm{O}_y},s^{\mathrm{O}_z})$,  
\begin{align}
\label{eq:S1}
%  S_1&=\frac{1}{\sqrt{3+\kappa_1^2}}(0,-\kappa_1,1,1,1)  \\
   \bar S_1&=\frac{1}{1+\kappa_1}(0,-\kappa_1,1,1,1)  \\
  \label{eq:S2}
%  S_2&=\frac{1}{\sqrt{4+\kappa_2^2}}(-\kappa_2,1,1,1,1) \\
    \bar S_2&=\frac{1}{1+\kappa_2}(-\kappa_2,1,1,1,1) \\
  \label{eq:S3}
%  S_3&=\sqrt{\frac{2}{3}}(0,0,-\frac{1}{2},-\frac{1}{2},1)
   \bar S_3&=\frac{2}{3}(0,0,-\frac{1}{2},-\frac{1}{2},1)
\end{align}
and shown in Figs.~\ref{fig:Si-sketch}(a)-(c). 
The coefficients $\kappa_1=\frac{3m^\mathrm{O}}{m^\mathrm{Ti}}$ and $\kappa_2=\frac{3m^\mathrm{O}+m^\mathrm{Ti}}{m^\mathrm{Sr}}$ ensure the center of mass is not displaced for any of the $\bar S_i$ modes. That is, $\sum_j m^j \bm{r}^j=0$ when summing over all the $j$ atoms with atomic mass $m^j$ and displacement $r^j=u_is^j$ in the unit cell for each of the $\bar S_i$ modes. The coefficient $u_i$ sets the amplitude of basis mode $i$  in the general displacement $\bm{ \bar U}$. The $\bar S_i$ modes in Eqs.~\eqref{eq:S1}-\eqref{eq:S3} have been normalized so that their amplitude $u_i$ is equal to the relative displacement of the two bodies in the mode. For instance, $u_1$ is the relative atomic displacement between Ti and the O cage in the $\bar S_1$ mode, $r^{\mathrm{O}}-r^\mathrm{Ti}=u_1(s_1^\mathrm{O}-s_1^\mathrm{Ti})=u_1$. Similarly, $u_2$ is the relative displacement between Sr and the Ti-O cage in mode $\bar S_2$. This normalization reduces the two-body problem into a one-body problem with a reduced mass when deriving the electron-phonon Hamiltonian, as will be shown in Section~\ref{sec:el-ph}.
Note the bar symbol indicates a vector spanned by the atoms of the unit cell (as in $\bar S_i$), whereas the vector referring to the Cartesian coordinates of the atomic displacements is specified by bold notation (as in $\bm{\hat{n}}_i$).

 As it is well known, the long-range Coulomb interaction partially lifts the three-fold degeneracy of polar modes into a high-energy longitudinal mode and low-energy doubly degenerate transverse modes~\cite{Cowley1964}. 
 Thus, the soft mode of STO is transverse and, 
in general, it is a linear combination of all three $\bar S_i$ modes. According to several studies~\cite{Cowley1964,1967Axe,Harada1970,Vogt1988}, its atomic displacements are close to the $\bar S_1$ mode [Eq.~\eqref{eq:S1}], also known as the Slater mode~\cite{slater}, where the Ti atom vibrates opposite to the O octahedron [see Fig.~\ref{fig:Si-sketch}(a)]. Because of the strong sensitivity of the electron-phonon coupling to the soft-mode eigenvector, we anticipate that even a small deviation from a pure Slater mode can have important consequences for superconductivity. 

Rigorously speaking, the above analysis in terms of three basis modes is only valid in the cubic phase. The presence of the AFD distortion requires an enlargement of the basis.
Indeed, below $105$ K, as the symmetry of STO is lowered to a tetragonal structure belonging to the $I4/mcm$ space group ($D_{4h}$ point group), the $T_{1u}$ polar mode of the cubic state splits into: (a) a $A_{2u}$ irrep with a polar axis along $[001]$ ($C_{4v}$) and (b) a $E_{u}$ irrep with a polar axis perpendicular to $[001]$. This split of the soft mode has been tracked in $T$ by hyper-Raman spectroscopy~\cite{Yamanaka2000}: $\hbar\omega_{E_u}\sim 1$ meV and $\hbar\omega_{A_{2u}}\sim 2$ meV at 7K. 

The analysis of polar modes for case (a) is simpler, as the basis of symmetry modes Eqs.~\eqref{eq:S1}-\eqref{eq:S3} for the $T_{1u}$ mode of $O_h$ in the high-$T$ cubic phase is also a complete basis for the $A_{2u}$ mode in the low-$T$ tetragonal phase. Therefore, in this case an enlargement of the basis is not needed. Of course, the atomic displacements are restricted along the tetragonal $z$ axis for this irrep, $\hat{n}_{i}^{A_{2u}}\parallel[001]$, leading to a polar tetragonal structure with $I4cm$ lower symmetry. Although here we will focus on the paraelectric phase, we note that
the out-of-plane polar mode has also been observed by electron microscopy and optical second harmonic generation in strained STO films in the symmetry broken polar phase ~\cite{Harter2019,salmani2020order,salmani2020polar,salmani2021role} highlighting the relevance of this symmetry. 

For case (b), in general, we notice that a distortion belonging to the $E_u$ irrep can lead to various lower-symmetry structures ($Ima2$, $Fmm2$ or $Cm$), all of which have a polar axis perpendicular to the tetragonal axis $[001]$. 
Here we will focus on $Ima2$ ($C_{2v}$), a space group with a polar axis parallel to the tetragonal in-plane axis, i.e. along the pseudocubic direction $[1\bar{1}0]$ (see Fig.~\ref{fig:Si-sketch}). This choice is justified by the fact that this mode has been experimentally reported in the ferroelectric phase of isotope O$_{18}$ and Ca substitution systems~\cite{shigenari2015raman,bednorz1984sr} and the optically excited metastable polar phase of STO~\cite{nova2019metastable}.
%MNG Add also Kozima \emph{ab initio}
Because the in-plane O atoms are not equivalent in $Ima2$, and Ti and in-plane O atoms are allowed to move orthogonally to the polar axis, the dimension of the basis symmetry modes has to be expanded to five. That is, besides the three $\bar S_i$ modes presented in Eqs.~\eqref{eq:S1}-\eqref{eq:S3}, with displacements along the polar axis (i.e. $\bm{\hat{n}}_i= \bm{\hat{n}}_p\parallel [1\bar{1}0]$ for $i=1,2,3$) one needs to add to the subspace another two $\bar S_i$ modes with an amplitude along the perpendicular direction (i.e. $\bm{\hat{n}}_i\parallel [110]\perp \bm{\hat{n}}_p$ for $i=4,5$) to obtain a complete basis,
\begin{align}
    \label{eq:S4}
    \bar S_4&=\frac{1}{2}(0,1,0,0,0),\\
    \label{eq:S5}
    \bar S_5&=\frac{1}{2}(0,0,1,-1,0).
\end{align}
These two modes are shown in Figs.~\ref{fig:Si-sketch}(d) and \ref{fig:Si-sketch}(e), respectively. 

In general, normal modes will not be made of collinear displacements. Still, they can be decomposed into the present basis, with each element representing collinear displacements.  Indeed, different elements of the expansion can have displacements in different directions although they contribute to the same polarization vector. 
%However, they can be decomposed into the present basis, in which each element is defined by collinear displacements. 

Table~\ref{tab:Si-modes} summarizes the polar distortion directions $\bm{\hat{n}}_i$ of the different $\bar S_i$ for both the $E_u$ ($Ima2$) and $A_{2u}$ ($I4cm$) modes we will consider throughout this work.

In general for a mode $i$ with displacement amplitude $u_i$ and associated with the polarization vector $\bm{\hat n}_p$, irrespectively of the direction of atomic displacements, we define its associated polarization vector  as 
\begin{align}
    \label{eq:ui1}
    \bm{u}_{i}=u_i \bm{\hat n}_p.
\end{align}
Notice that formally $\bm{u}_{i}$ should be multiplied by an effective  charge to be a real polarization. On the other hand, such charge does not play any role in the present context and will be omitted. 

Near the zone center but for finite $\bm{q}$ it is important to consider the long-range Coulomb interaction which will split transverse and longitudinal modes. In this case we will restrict to the symmetrized modes $i=1,2,3$ and we will assume that for $\bm{q}$ along an arbitrary direction we can decompose mode $i$ into the set of the same $i$
modes along directions defined by ${\bm q}$ as done in Ref.~\cite{gastiasoro2022theory} for the Slater mode. For simplicity, deviations of the polarization vector from pure longitudinal or transverse directions due to non-spherical symmetry will be neglected. In this way 
we can generalize Eq.~\eqref{eq:ui1} to 
\begin{align}
    \label{eq:ui2}
    \bm{u}_{i}(\bm{q})=u_i (\bm{q})\bm{\hat n}_p(\bm{q}),
\end{align}
which will be used next to discuss the general interaction with polar modes.  
%We can also define the displacement associated with a basis mode as
%\begin{align}
%    \label{eq:rj}
%    \bm{u}^{j}(\bm{q})=u_i(\bm{q})s^j\bm{\hat n}_i, 
%\end{align}
%which, as discussed, need not be collinear with the polarization. 

\begin{table}
 \caption{Symmetry coordinates of polar $E_u$ and $A_{2u}$ modes for a general distortion $\bm{\bar U}=\sum_{i}u_i\bm{\hat{n}}_i\bar S_i$ in the tetragonal phase of STO. The atomic displacement coordinates $\bar S_i$ are given by Eqs.~\eqref{eq:S1}-\eqref{eq:S5}. The polar axis of the modes shown here is along the in-plane direction, $\bm{\hat{n}}^{E_u}_{p}\parallel [1\bar{1}0],[110]$, for $E_u$ modes and out-of-plane $\bm{\hat{n}}^{A_{2u}}_{p}\parallel[001]$ for $A_{2u}$ modes.  The symmetry modes analysis has been performed using the ISODISTORT tool~\cite{isodistort1, isodistort2}. }
\begin{ruledtabular}
     \setlength{\extrarowheight}{3pt}%
     \begin{tabular}{lccl}
        Irrep (space group, point group)&$\bm{\hat{n}}_p$ & $\bm{\hat{n}}_i$  & $\bar S_i$ \\
     \hline
        $E_u$ ($Ima2$, $C_{2v}$) & $[1\bar{1}0]$&  $[1\bar{1}0]$&  $\bar S_1$, $\bar S_2$, $\bar S_3$ \\
        & $[1\bar{1}0]$& $[110]$ &  $\bar S_4$, $\bar S_5$  \\
        & $[110]$&  $[110]$&  $\bar S_1$, $\bar S_2$, $\bar S_3$ \\
         & $[110]$&$[1\bar{1}0]$ & $\bar S_4$, $\bar S_5$  \\
    \hline 
        $A_{2u}$ ($I4cm$, $C_{4v}$) & $[001]$  & $[001]$&  $\bar S_1$, $\bar S_2$, $\bar S_3$ 
    \end{tabular}
\end{ruledtabular}
    \label{tab:Si-modes}
\end{table}

\section{Linear Rashba-like coupling}
\label{sec:linear}
\subsection{\emph{Ab initio} computation of couplings}
\label{sec:lineardft}

Having established the electronic structure of STO in Sec.~\ref{sec:electronic} and the relevant polar phonon modes around the zone center in Sec.~\ref{sec:Si}, in this section we proceed to study and estimate their coupling to linear order. 
In particular, we will show how a symmetry allowed linear coupling to transverse TO modes emerges naturally from induced hopping channels, and estimate the corresponding coupling constant for all electronic bands with the aid of \emph{ab initio} frozen-phonon results in tetragonal STO.

The linear coupling Hamiltonian between a polar distortion and electronic 
bands %with spinor $\psi^\dagger_n=(c^{\dagger}_{n+},c^{\dagger}_{n-})$ for band $n$
[Eqs.~\eqref{eq:tetbasis1}-\eqref{eq:Hband}] can be expressed as
\begin{equation}
 \mc{H}_u= \sum_{n n' \bm{k} \bm{q},\bar S_i }  \psi_n^\dagger(\bm{k}+\frac{\bm{q}}{2}) \bm{\Lambda}^{\bar S_i}_{nn'}(\bm{k},\bm{q}) \psi_{n'}(\bm{k}-\frac{\bm{q}}{2})  
 \label{eq:Husoc}
\end{equation}
with the coupling $ 2\times 2$ matrix $\bm{\Lambda}^{\bar S_i}_{nn'}(\bm{k},\bm{q}) $ in pseudospin space for a polar mode $\bar S_i$. 
The intra-band ($n'=n$) coupling matrix has the following form to linear order in $\bm{k}$ and $u_\alpha $ in a $D_{4h}$ point group:
\begin{eqnarray}
  \label{eq:Gamman}
     \bm{\Lambda}^{\bar S_i}_{nn}(\bm{k},\bm{q})=&&  k a\varepsilon_{\alpha\beta\gamma} u_\alpha \hat{k}_\beta \sigma_\gamma\left(\delta_{\alpha z}\tau^{\bar S_i}_{n,C}+\delta_{\beta z}\tau^{\bar S_i}_{n,B}+\delta_{\gamma z}\tau^{\bar S_i}_{n,A}\right)\nonumber \\
     =&& k a\Bigl[ \tau^{\bar S_i}_{n,A} (\hat k_y,-\hat k_x,0)\sigma_z 
     +\tau^{\bar S_i}_{n,B}\hat{k}_z(-\sigma_y,\sigma_x,0) \nonumber\\  &&+\tau^{\bar S_i}_{n,C}(0,0,\hat k_x\sigma_y-\hat k_y\sigma_x) \Bigr] \cdot\bm{u}_{i}(\bm{q}),
\end{eqnarray}
where $a$ is the lattice constant, $\varepsilon_{\alpha\beta\gamma}$ the Levi-Civita symbol, $\delta_{\alpha\beta}$ the Kronecker's delta-function, $\hat{k}_i=\bm{\hat{k}}\cdot \bm{\hat\imath}$ the Cartesian projection of the unitary momentum vector, and $\sigma_j$ the Pauli matrices for the pseudospin of the electronic bands. We also defined the couplings $\tau^{\bar S_i}_{n,l}$ with symmetry allowed irrep labels $l=A,B,C$.

Equation~\eqref{eq:Gamman} describes a Rashba-like linear-in-$k$ coupling between a polar distortion $\bm{u}_{i}(\bm{q})=u_i(\bm{q})\bm{\hat n}_p(\bm{q})$ of mode $\bar S_i$ and the electronic band $n$ with pseudospin $\bm\sigma$. 
The first form in Eq.~\eqref{eq:Gamman} makes evident that $z$ is a privileged axis in this structure and clarifies the meaning of the $\tau^{\bar S_i}_{n,l}$ coefficients which are associated with one member of the  triad $(\bm{u}_i,\bm k ,\bm\sigma)$ having a projection on the $z$-direction.

In the following we drop the index $\bar S_i$ from the couplings $\tau_{n,l}$ for simplicity, but emphasize that these couplings vary a lot from mode to mode.
The Rashba matrix Eq.~\eqref{eq:Gamman} has the most general form allowed by the symmetry of our tetragonal system to linear order in $\bm{k}$. It consists of couplings $\tau_{n,A}$ and $\tau_{n,B}$ ($\tau_{n,C}$) which couple to the corresponding $E_u$ ($A_{2u}$) modes with polar axis $\bm{\hat n}_p$ in the $xy$ plane (along the $z$ axis). These parameters are not related by symmetry, and we will estimate them using \emph{ab initio} computations in the following.
Note that in higher cubic symmetry Eq.~\eqref{eq:Gamman} simplifies to $k a[\tau^{\bar S_i}_n\bm{\hat{k}}\times\bm{\sigma} ]\cdot \bm{u}(\bm{q})$~\cite{Kozii2015,Gastiasoro2020,Sumita2020}.
  
%Comment: Ref.~\cite{Sumita2020} chose $\tau'_{n,b}=\pm \tau'_{n,a}$, which is not imposed by symmetry.
The interband coupling matrices ($n'\neq n$) in Eq.~\eqref{eq:Husoc} have a similar $k$-linear Rashba form. However, because we are considering a long-wavelength $\bm q\rightarrow 0$ phonon, the inter-band terms result in $k$-cubic Rashba intra-band terms upon perturbation. We therefore focus solely on intra-band terms with $n'=n$ [Eq.~\eqref{eq:Gamman}] in this work, which involve $k$-linear Rashba terms which can be directly extracted from \emph{ab initio} frozen-phonon computations. 

Finite Rashba couplings $\tau_{n,l}$ in Eq.~\eqref{eq:Gamman} cause the characteristic linear-in-$k$ band splitting $E_{n+}(\bm{k})-E_{n-}(\bm{k})\equiv \delta E_n(\bm{k})$. From the Rashba coupling matrix Eq.~\eqref{eq:Gamman} the splitting for band $n$ is generally given by the following expression
\begin{widetext}
\begin{align}
\label{eq:deltaEn}
   \delta E_n & (\bm{k},\bm{u}_{i}(\bm{q}))=2 k a u_{i}(\bm{q})\times\nonumber\\
   &\times\sqrt{\tau_{n,A}^{2}\left(\hat{k}_y\hat{n}_{px}-\hat{k}_x\hat{n}_{py}\right)^2+\tau_{n,B}^{2}\hat{k}_z^2\left(\hat{n}_{px}^2+\hat{n}_{py}^2\right)+\tau_{n,C}^{2}\left(\hat{k}_x^2+\hat{k}_y^2\right)\hat{n}_{pz}^2 -2\tau_{n,B}\tau_{n,C}\left(\hat{k}_x\hat{n}_{px}+\hat{k}_y\hat{n}_{py}\right)\hat{k}_z\hat{n}_{pz}},
\end{align}
\end{widetext}
which peaks (vanishes) along momenta $\bm k$ perpendicular (parallel) to the polar axis $\bm{\hat{n}}_p$ of the mode.
As we show in the following, one can use this band split to extract the  Rashba couplings of each band $\tau_{n,l}$ to each mode $\bar S_i$. 
For this purpose, we need to particularize Eq.~\eqref{eq:deltaEn} for the direction given by the polar axis of the polar modes. We obtain 
%Obtaining the analogous expression for the $k$-linear Rashba model Eq.~\eqref{eq:deltaEn}, we get 
the following pseudospin split for a mode $\bar S_i$ 
\begin{align}
    \label{eq:deltaEnuy}
    \frac{\delta E_n (\bm{k},u_{i}[1\bar{1}0])}{2u_{i}}&= k a \sqrt{\frac{\tau_{n,A}^{2}}{2}(\hat{k}_y+\hat{k}_x)^2+\tau_{n,B}^{2}\hat{k}_z^2} \\
    \label{eq:deltaEnuz}
     \frac{\delta E_n (\bm{k},u_{i}[001])}{2 u_i}&= k a\tau_{n,C}\sqrt{\hat{k}_x^2+\hat{k}_y^2}
\end{align}
with an amplitude $u_i$ and a polar axis $\bm{\hat{n}}_p\parallel[1\bar{1}0]$ and $\bm{\hat{n}}_p\parallel[001]$, respectively.

To obtain the couplings $\tau_{n,l}$, we computed by first principles the reconstructed electronic band structure of tetragonal STO in the presence of a frozen phonon $\bm{u}_{i}(\bm{q}=0)$ for all $\bar S_i$ modes listed in Table~\ref{tab:Si-modes} for both $E_u$ and $A_{2u}$ irreps. 

\begin{figure}
    \centering
    \includegraphics[width=\linewidth]{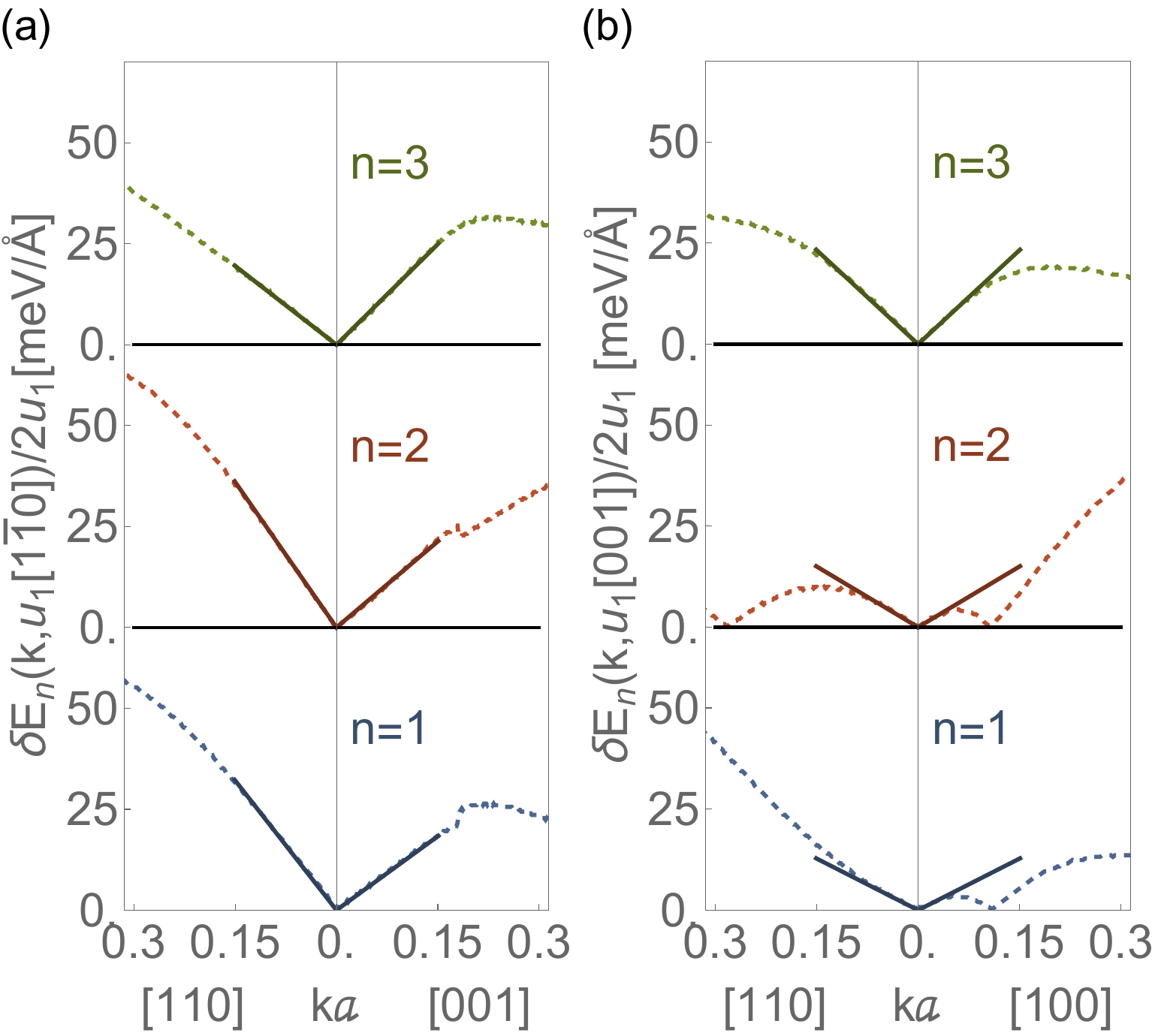}
    \caption{Electronic band split $\delta E_n\left(\bm{k},u_{1}(q=0)\bm{\hat{n}}_{p}\right)$ in the presence of a frozen Slater polar mode [Eq.~\eqref{eq:S1}] normalized by twice the amplitude $2 u_{1}$ along the specified $\bm{k}$ directions for the three bands of tetragonal STO [Fig.~\ref{fig:bands}(b)]. The polar axis of the mode is along (a) $\bm{\hat{n}}_{p}^{E_u}\parallel[1\bar{1}0]$ and (b) $\bm{\hat{n}}_p^{A_{2u}}\parallel[001]$ in pseudocubic coordinates. Dashed lines are \emph{ab initio} results from frozen-phonon distortions. Full lines are the $k$-linear Rashba split model Eq.~\eqref{eq:deltaEnuy} in (a) and Eq.\eqref{eq:deltaEnuz} in (b), up to $ka=0.15$. All shown $\bm{k}$ directions are perpendicular to the polar axis $\bm{\hat{n}}_p$.
    }
    \label{fig:splitdft}
\end{figure}

Frozen-phonon distorted structures have been constructed by displacing atoms along each symmetrized mode while keeping fixed the optimized lattice parameters. In order to identify the linear regime in $\bm u_i$, band-structure calculations have been performed for several values of the displacement amplitude up to 0.1~\AA.
By fitting the linear-$k$ regime of the DFT band-splitting in Fig.~\ref{fig:splitdft} along various momentum directions, we can obtain for each band $n$ the $E_u$ Rashba couplings $\tau_{n,A}$ and $\tau_{n,B}$, and the $A_{2u}$ coupling $\tau_{n,C}$.
The sign of these couplings is extracted from the averaged spin-polarization values of each band obtained from the same \emph{ab initio} computation. Keeping track of these signs is very important when computing the coupling to a mode that is combination of the $\bar S_i$ modes, as will become clear below.

As an example, we show the results of the frozen Slater mode $\bar S_1$ [Eq.~\eqref{eq:S1} and Fig.~\ref{fig:Si-sketch}(a)].
The resulting band split of each $n$ band $\delta E_n(\bm k)$ found by \emph{ab initio} is shown by the dashed lines in Fig.~\ref{fig:splitdft}(a) and Fig.~\ref{fig:splitdft}(b), for $\bm{u}_{1}$ with a polar axis along the in-plane $\bm{\hat{n}}_p\parallel[1\bar{1}0]$ ($E_u$ mode) and out-of-plane $\bm{\hat{n}}_p\parallel[001]$ ($A_{2u}$ mode) pseudocubic directions, respectively.
%For a discussion on the experimental relevance of these two polar distortions see Section~\ref{sec:Si}.
As seen in Fig.~\ref{fig:splitdft}, for small enough momenta all bands show a linear-$k$ split (full lines), but the momentum amplitude beyond which deviations of $k$-linearity become significant depends on the band, the polar axis and the direction of momentum. 
In fact, while for the $E_u$ mode the split is robustly linear around the zone center with small deviations beyond $ka\sim 0.2$, for the $A_{2u}$ mode strong non-linear features appear already at small $ka \sim 0.05$ for the two lowest bands. This highlights the limitation of a conventional Rashba linear-$k$ model Eq.~\eqref{eq:Gamman} to describe the coupling between the bands and some of the polar modes in this system. We will come back to this important point in Section~\ref{sec:SC}.

\begin{figure}
    \includegraphics[width=\linewidth]{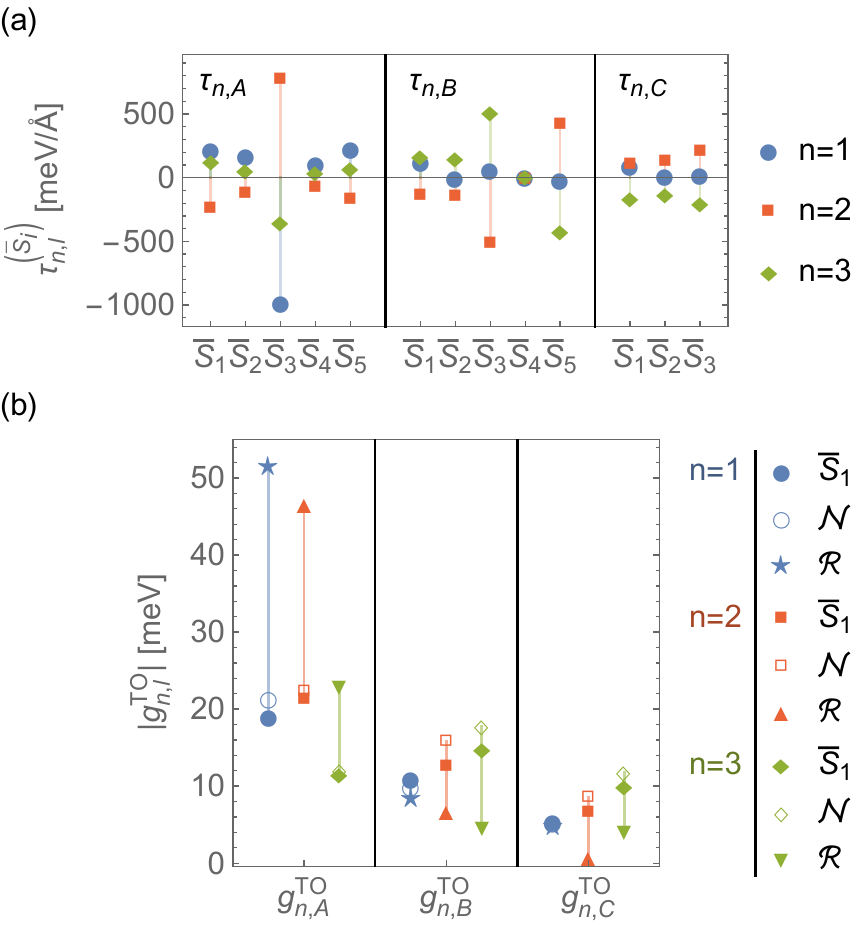}
    \caption{ (a) $k$-linear Rashba couplings $\tau_{n,A}$, $\tau_{n,B}$ and $\tau_{n,C}$ in Eq.~\eqref{eq:Gamman} in meV/$\AA$ for band $n$ and mode $\bar S_i$ [Eqs.~\eqref{eq:S1}-\eqref{eq:S5}]  in tetragonal STO. They were obtained from fits of the spin-split Eqs~\eqref{eq:deltaEnuy}-\eqref{eq:deltaEnuz} to frozen phonon \emph{ab initio} results. The numerical values are listed in~\cite{SM}.
    (b) Absolute value of the estimated electron-FE-mode Rashba couplings [Eq.~\eqref{eq:galpha}] for band $n$ for three possible eigenvectors of the FE mode: a pure $\bar S_1$ mode, eigenvector from neutron data [Eq.~\eqref{eq:neutron}] and from Raman data [Eq.~\eqref{eq:raman}]. We have used experimental optical gaps
    $\hbar\omega^{E_u}_\mathrm{TO}=1$ meV and $\hbar\omega^{A_{2u}}_\mathrm{TO}=2$ meV, and electronic momentum $k_F a=0.3$.}
    \label{fig:lambda}
\end{figure}

The Rashba couplings to the rest of the $\bar S_i$ modes belonging to $E_u$ and $A_{2u}$ irreps (see Table~\ref{tab:Si-modes}) have been also estimated by the same fitting procedure to frozen phonon \emph{ab initio} computations; they are shown together with those of the Slater mode in Fig.~\ref{fig:lambda}(a) (also listed in~\cite{SM}). 
As seen, for all $\bar S_i$ modes the $E_u$ coupling $\tau_{n,A}$ is larger in the lowest two bands ($n=1,2$) than the highest band ($n=3$). This hierarchy is reversed for the couplings $\tau_{n,B}$ and $\tau_{n,C}$ where the highest two bands ($n=2,3$) show larger couplings than the lowest band ($n=1$). 
Remarkably, the Rashba coupling to the $\bar S_3$ mode, with apical oxygen  atoms O$_z$ moving opposite to in-plane O atoms O$_{x,y}$ distorting the octahedra [see Eq.~\eqref{eq:S3} and Fig.~\ref{fig:Si-sketch}(c)] can be an order of magnitude larger than the other couplings.  As we will show in the following section, this gigantic Rashba coupling has important consequences for the electron coupling to the soft mode. Indeed, an enlarged electron-phonon coupling follows from a modest contribution of $\bar S_3$ to any polar mode.  

\subsection{Real-space origin of the coupling}

\begin{figure}
    \centering
    \includegraphics[width=\linewidth]{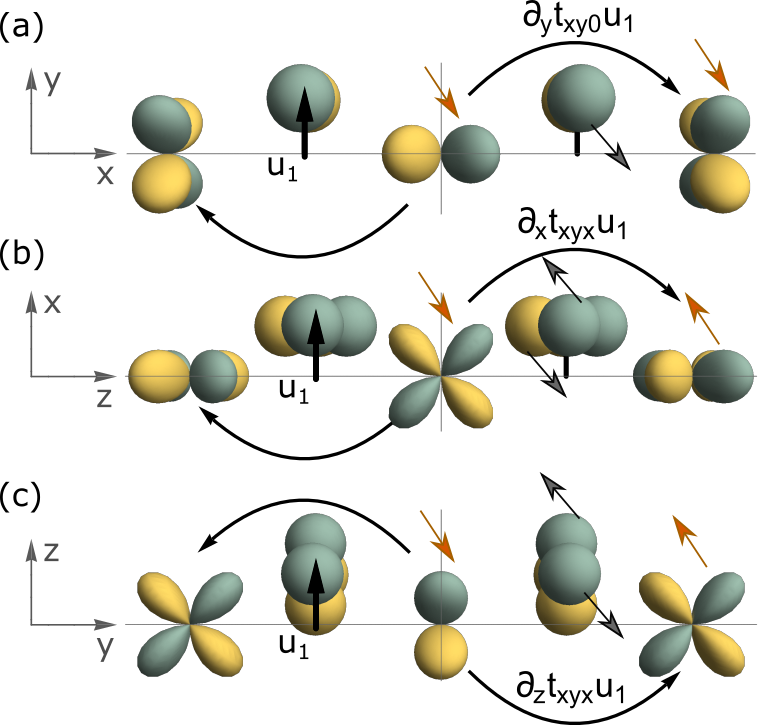}
    \caption{Induced inter-orbital hopping elements in Eq.~\eqref{eq:tauxy} between $d$-orbital $\nu=y$ with spin $\downarrow$ (orange arrow) at the origin and $d$-orbital $\mu=x$ with spin $s$ (orange arrow) at neighboring Ti atoms, mediated by the $p$-orbitals with spin $s'$ (gray arrow) in the bridging oxygens. (a) Spin-conserving process $\partial_y t_{xy0}u_1$ with $s=\downarrow$ mediated by $|p_z\downarrow>$. 
    Spin-flip processes $s=\uparrow$ when allowing for SOC of the oxygen $\xi_O$ for (b)  $\partial_x t_{xyx}u_1$ mediated by $|p_z\downarrow>$ and $|p_y\uparrow>$  with matrix-element $i\xi_O$, and (c)  $\partial_z t_{xyx}u_1$ mediated by $|p_x\downarrow>$ and $|p_z\uparrow>$ with matrix-element $\xi_O$. 
    All three hopping elements change sign along the horizontal bond (black curved arrows) with a finite polar Slater displacement $\bm{u}_1$ along the vertical axis, specified in each panel. 
    The spins on the negative horizontal axis, not shown for clarity, are mirror symmetric with respect to those shown on the positive side.}
    \label{fig:hopping}
\end{figure}

We will now show how the symmetry allowed coupling Eq.~\eqref{eq:Gamman} emerges when considering microscopic processes in real space.
In the presence of a polar distortion $\bm{\bar U}=\sum_{i}u_i\bm{\hat{n}}_i\bar S_i$ of the lattice, new terms are allowed in the Hamiltonian Eq.~\eqref{eq:tb-model} for $t_{2g}$ electrons around the zone center. These new terms include effects such as the polarization of the orbitals and induced hopping channels which are symmetry forbidden in the absence of the distortion~\cite{Petersen2000,Khalsa2013,Zhong2013,Djani2019,gastiasoro2022theory,kumar2022spin}. 
We thus consider the following Hamiltonian with the spinor of the $t_{2g}$ orbitals $\psi^\dagger_\mu=(c^{\dagger}_{\mu+},c^{\dagger}_{\mu-})$, 
\begin{equation}
    \mc{H}_u= \sum_{\bm{k} \bm{q} \mu \nu j}  \psi^\dagger_{\mu}(\bm{k}+\frac{\bm{q}}{2})t_{\mu\nu j}(\bm{k},\bm{q})\sigma_j \psi_{\nu}(\bm{k}-\frac{\bm{q}}{2})+\mathrm{h.c.}
    \label{eq:Hu}
\end{equation}
which describes the induced hopping between a $d$-orbital $\mu$ and a nearest neighbor $d$-orbital $\nu$. The Pauli matrices $\sigma_j$ represent spin-independent ($j=0$) as well as spin-dependent ($j=x,y,z$) hopping processes.  Fig.~\ref{fig:hopping} shows some examples for $j=0,x$ and $y$. 
Around the zone center ($\bm{k}\rightarrow 0$, $\bm{q}\rightarrow 0$), the new allowed terms have the following form
\begin{equation}
     t_{\mu\nu j}(\bm{k},\bm{q})\approx \sum_{ilm}\frac{\partial t_{\mu\nu j}}{\partial u_{i,l}(\bm{q})}u_{i,l}(\bm{q})  k_m  a  
     \label{eq:taumunu}
\end{equation}
to linear order in the polar distortion $\bm{u}_{i}(\bm{q})=u_{i}(\bm{q})\bm{\hat{n}}_{p}(\bm{q})$  for a mode $\bar S_i$ with amplitude $u_{i,l}(\bm{q})$ in Cartesian coordinate $l$, and polar axis $\bm{\hat{n}}_{p}$. 
In the case of tetragonal STO, as discussed in the previous section, the polar axis we are considering are the in-plane $\bm{\hat{n}}^{E_u}_{p}\parallel [1\bar{1}0]$ for $E_u$ modes and out-of-plane $\bm{\hat{n}}^{A_{2u}}_{p}\parallel [001]$ for $A_{2u}$ modes in pseudocubic coordinates.
In general, the precise form of the terms allowed in Eq.~\eqref{eq:taumunu} is set by symmetry; and thus it depends on the pair of orbitals $\mu$ and $\nu$ involved, as well as the direction of the polar axis $\bm{\hat{n}}_{p}$ associated with the distortion $\bm{\bar U}$. 

As a concrete example, we explicitly consider in the following the case of $\mu=x$ and $\nu=y$ orbitals. The lowest electronic band in the tetragonal state ($n=1$) is formed by only these two orbitals at the zone center [see Eq.~\eqref{eq:tetbasis1}], and thus induced hopping amplitudes involving these two orbitals are the relevant terms for the coupling of the lowest band to polar modes. 
For a general polarization vector defined by Eq.~\eqref{eq:ui2}, for mode $\bar S_i$, the following inter-orbital hopping elements are allowed in Eq.~\eqref{eq:Hu}: 
\begin{align}
\sum_jt_{xy j}(\bm{k},\bm{q})\sigma_j&\approx 2i\partial_x t_{xy0} u_i(\bm{q})k a
\bigl[ -\hat{k}_y\hat{n}_{px}+\hat{k}_x\hat{n}_{py}\bigr] \sigma_0\nonumber \\
&+ 2\partial_x t_{xyx} u_i(\bm{q})k a\bigl[ - \hat{n}_{px}\sigma_x+ \hat{n}_{py}\sigma_y \bigr] \hat{k}_z\nonumber \\
&+ 2\partial_z t_{xyx} u_i(\bm{q})  k a\bigl[ -\hat{k}_x \sigma_x +\hat{k}_y  \sigma_y \bigr]\hat{n}_{pz}
\label{eq:tauxy}
\end{align}
where we have used the shorthand notation $\frac{\partial t_{\mu\nu j}}{\partial u_{i,l}(\bm{q})}\equiv \partial_l t_{\mu\nu j}$.
The first term in Eq.~\eqref{eq:tauxy} corresponds to a spin-conserving ($\sigma_0$) hopping channel with amplitude $\partial_x t_{xy0}$ which changes sign with hopping direction, shown in Fig.~\ref{fig:hopping}(a). It couples only to the in-plane components of the polar distortion axis $\hat{n}_{px}$ and $\hat{n}_{py}$. The second and third terms describe spin-flip ($\sigma_{x,y}$) hopping processes instead, and couple to the in-plane component of the polar axis, through the hopping amplitude $\partial_x t_{xyx}$ [Fig.~\ref{fig:hopping}(b)], as well as to the out-of-plane component $\hat{n}_{pz}$ through the hopping amplitude $\partial_z t_{xyx}$ [Fig.~\ref{fig:hopping}(c)]. Microscopically one obtains spin-flip hopping terms by extending SOC to the bridging $p$-orbitals of the oxygen atom [as shown in Figs.~\ref{fig:hopping}(b)-(c)], or by considering virtual processes between the $t_{2g}$ and $e_g$ manifolds~\cite{Djani2019}. 
Each $\bar S_i$ mode will generally induce different hopping amplitudes, which then result in different couplings to the electrons. 
Other $t_{\mu\nu j}(\bm{k},\bm{q})$ induced hopping elements between different pairs of $t_{2g}$ orbitals in Eq.~\eqref{eq:Hu} can be similarly obtained. 

We can now connect the symmetry allowed couplings in Eq.~\eqref{eq:Gamman}, to these microscopic processes by projecting Eq.~\eqref{eq:Hu} to the band basis [Eqs.~\eqref{eq:tetbasis1}-\eqref{eq:tetbasis3}] of the non-interacting electrons.
In general each coupling element $\tau_{n,l}$ is a linear combination of the different induced hopping derivatives $\partial_l t_{\mu\nu j}$ allowed in Eq.~\eqref{eq:taumunu}.
For instance, the induced hopping derivatives between the $x$ and $y$ orbitals considered in Eq.~\eqref{eq:tauxy}, are connected to the Rashba couplings for the lowest band $n=1$ [Eq.~\eqref{eq:tetbasis1}] in the following way,
\begin{align}
\label{eq:lambdan1}
    \tau_{1,A} =-2\partial_x t_{xy0}; \ \tau_{1,B} =2\partial_x t_{xyx}; \ \tau_{1,C} =-2\partial_z t_{xyx}.
\end{align}
Similar expressions for the other two electronic bands $n=2,3$ can also be derived following the same procedure. These expressions are more involved than the particularly simple expressions in Eq.~\eqref{eq:lambdan1} for $n=1$.

To close this section we comment on the importance of  the different terms in [Eq.~\eqref{eq:lambdan1}], relevant for the coupling to the lowest band $n=1$. %[Eq.~\eqref{eq:tetbasis1}]. 
We see from Fig.~\ref{fig:lambda}(a)
%the values we obtained from \emph{ab initio}
%listed in Table~\ref{tab:couplings} 
that for $n=1$ and for all modes $\bar S_i$ except Slater ($\bar S_1$), $\tau_{1,A}\gg  \tau_{1,B},\tau_{1,C}$, implying that one can safely neglect the spin-flip induced hopping terms $\partial_x t_{xyx}$ and $\partial_z t_{xy}$ in Eq.~\eqref{eq:tauxy}.
For a pure Slater mode $\bar S_1$ mode, however, $\tau_{1,A}$ is  of the same order of magnitude as $\tau_{1,B}$ and $\tau_{1,C}$, hence 
the spin-flip processes (and the symmetry equivalent virtual processes to the $e_g$ manifold) are in principle not negligible.
However, since the $\bar S_3$ spin-conserving coupling $\tau_{1,A}^{(\bar S_3)}$ is very large, a small admixture of this mode allows to neglect spin-flip processes when considering the linear coupling between the lowest band and the polar mode.

\section{Electron-polar-phonon Hamiltonian}
\label{sec:el-ph}

In order to obtain an electron-phonon Hamiltonian, we quantize the general atomic displacements of Eq.~\eqref{eq:U} 
%in the polar mode amplitude $\bm{u}_{i}(\bm{q})$ appearing in Eq.~\eqref{eq:Gamman} and
by decomposing them into a set of normal modes $\alpha$: 
\begin{equation}
\label{eq:utoA}
 \bm{r}^{j}(\bm{q})=\sum_\alpha \frac{\bm{e}^j_\alpha(\bm{q})}{\sqrt{m^{j}}}\sqrt{\frac{\hbar}{2\mathcal{N}\omega_{\bm{q}\alpha}}}\mathcal{\hat{A}}_{\bm{q}\alpha}    
\end{equation}
Here $\mathcal{N}$ is the number of unit cells, $\mathcal{\hat{A}}_{\bm{q}\alpha}=\hat{a}_{\bm{q}\alpha}+\hat{a}^\dagger_{-\bm{q}\alpha}$ is the phonon operator of mode $\alpha$ with frequency $\omega_{\bm{q}\alpha}$ and normalized eigenvector $\bm{e}^j_\alpha(\bm{q})$. To proceed, we need an analogous expression for the polarization amplitude $u_i$ appearing in Eqs.~\eqref{eq:ui2}-\eqref{eq:Gamman}. We write the ansatz,
$$u_i=\sum_{\alpha} \frac{\varphi_{\alpha i}}{\sqrt{ m_u}} \sqrt{\frac{\hbar}{2\mathcal{N}\omega_{\bm{q}\alpha}}}\mathcal{\hat{A}}_{\bm{q}\alpha} $$
with $m_u$ the atomic mass constant. Inserting it in Eq.~\eqref{eq:U} one obtains that the coefficients $\varphi_{\alpha i}$ are determined by the decomposition of the normalized displacement vector in the complete basis of the $\bar S_i$ modes [Eqs.~\eqref{eq:S1}-\eqref{eq:S5}]: 
\begin{align}
\label{eq:expansionphi}
    \sqrt{m_u}\left(\frac{\bm{e}^\mathrm{Sr}_\alpha}{\sqrt{m^{\mathrm{Sr}}}},\frac{\bm{e}^\mathrm{Ti}_\alpha}{\sqrt{m^{\mathrm{Ti}}}},\frac{\bm{e}^\mathrm{O_x}_\alpha}{\sqrt{m^{\mathrm{O}}}},\frac{\bm{e}^\mathrm{O_y}_\alpha}{\sqrt{m^{\mathrm{O}}}},\frac{\bm{e}^\mathrm{O_z}_\alpha}{\sqrt{m^{\mathrm{O}}}}\right)\nonumber\\=\sum_i \varphi_{\alpha i} \hat{\bm{n}}_i\bar S_i
    \end{align}
with $\varphi_{\alpha i}$ set to normalize the polarization eigenvector of the $\alpha$ mode, i.e. $\sum_j |\bm{e}^j_\alpha|^2=1$.
As mentioned before, for a given mode $\alpha$, displacements do not need to be collinear even though they concur to the same polar axis $\bm{\hat{n}}_p^{\alpha}$.

With the above quantization, we obtain the following electron-polar-phonon Hamiltonian for a mode $\alpha$
\begin{equation*}
     \mc{H}_u= \frac{1}{\sqrt{\mathcal{N}}}\sum_{n \bm{k} \bm{q}\alpha }  \psi_n^\dagger(\bm{k}+\frac{\bm{q}}{2}) g^\alpha_{n}(\bm{k},\bm{q}) \psi_{n}(\bm{k}-\frac{\bm{q}}{2})\mathcal{\hat{A}}_{\bm{q}\alpha} 
\end{equation*}
with coupling function
\begin{align}
\label{eq:galphakq}
     g^{\alpha}_{n}(\bm{k},\bm{q}) &= 
     %\sqrt{\frac{\hbar}{2M_{S_i}\omega_{\bm{q}\alpha}}} \vartheta_{S_i} k a\times\nonumber \\ 
     %&\Bigl[\lambda'_{n,a} (-\hat{k}_y,\hat{k}_x,0)\sigma_z 
     %+\lambda'_{n,b} \hat{k}_z(-\sigma_y,\sigma_x,0) \nonumber \\
     %&+\lambda'_{n,c}(0,0,-\hat{k}_y\sigma_x+\hat{k}_x \sigma_y)\Bigr]\cdot \bm{\hat{n}}_{\alpha}(\bm{q})\\ &\equiv 
     \Bigl[g^{\bm{q}\alpha}_{n,A} (\hat{k}_y,-\hat{k}_x,0)\sigma_z 
     +g^{\bm{q}\alpha}_{n,B} \hat{k}_z(-\sigma_y,\sigma_x,0) \nonumber \\
     &+g^{\bm{q}\alpha}_{n,C}(0,0,\hat{k}_x\sigma_y-\hat{k}_y \sigma_x)\Bigr]\cdot \bm{\hat{n}}_p^{\alpha}(\bm{q}). 
\end{align}
For each $\alpha$ mode we have defined the 
%unit polarization vector $\bm{\hat{n}}_{\alpha}(\bm{q})$ and
electron-phonon matrix element:
\begin{equation}
\label{eq:galpha}
    g^{\bm{q}\alpha}_{n,l}=  k a \sqrt{\frac{\hbar}{2m_{u}\omega_{\bm{q}\alpha}}}\sum_i\varphi_{\alpha i}\tau^{(\bar S_i)}_{n,l}\equiv k a \sqrt{\frac{\hbar}{2m_{u}\omega_{\bm{q}\alpha}}}\tau^{(\alpha)}_{n,l}
\end{equation}
where $l=A,B,C$.
%We will also take the long-wavelength limit $\omega_{\bm{q}\alpha}\approx \omega_\alpha$ when considering the coupling to zone-center polar modes. 
Equation~\eqref{eq:galpha} shows that the $k$-linear Rashba coupling $\tau^{(\alpha)}_{n,l}$ of mode $\alpha$ is a weighted sum of the Rashba couplings $\tau^{(\bar S_i)}_{n,l}$ [shown in Fig.~\ref{fig:lambda}(a)] with the coefficients $\varphi_{\alpha i}$ weighing the contribution of the  $\bar S_i$ modes to the normal mode $\alpha$. 

Because of the strongly anharmonic nature of the problem~\cite{zhou2018electron,he2020anharmonic,shin2021quantum,fauque2022mesoscopic}, the eigenvector $\bm{e}^j_\alpha(\bm{q})$ [Eq.~\eqref{eq:expansionphi}] of the soft mode is particularly difficult to determine accurately, both theoretically and experimentally. 
Eqs.~(\ref{eq:utoA})-(\ref{eq:galpha}) allow to  compute the coupling to polar modes with arbitrary eigenvectors so they can be used to determine, for example, the coupling to the soft polar mode from better refined eigenvectors in future studies. In other words, we have separated the problem of determining the coupling to the soft mode from the problem of determining its eigenvector.

We illustrate the evaluation of the electron-phonon matrix elements in Eq.~\eqref{eq:galpha} by assuming first the $\alpha$ mode to be a pure Slater mode $\bar S_1$ [Eq.~\eqref{eq:S1}]. 
Then the only non-zero coefficient of the expansion in Eq.~\eqref{eq:expansionphi} is $\varphi_{\alpha 1}=\sqrt{\frac{m_u}{\mu_{\bar S_1}}}=0.204$, where we have introduced the reduced mass of the Slater mode $\mu^{-1}_{\bar S_1}=\left(m^\mathrm{Ti}\right)^{-1}+\left(3m^\mathrm{O}\right)^{-1}$. In this case the coupling Eq.~\eqref{eq:galpha} is then reduced to the following simple expression~\cite{gastiasoro2022theory}
\begin{equation}
\label{eq:gS1}
     g^{\bm{q}\alpha}_{n,l}= k a \sqrt{\frac{\hbar}{2\mu_{\bar S_1}\omega_{\bm{q}\alpha}}}\tau^{( \bar S_1)}_{n,l}.
\end{equation}
Substituting the estimated\cite{SM} Rashba couplings $\tau^{( \bar S_1)}_{n,l}$ for the Slater mode $\bar S_1$ and the experimental zone center frequency of the soft FE mode $\omega_{\bm{q}\alpha}=\omega_\mathrm{TO}^{E_u}$ and $\omega_\mathrm{TO}^{A_{2u}}$ we obtain the electron-TO couplings for STO listed in Table~\ref{tab:gcoupling} (under $\bar S_1$ columns). The value for $l=A$ in the first column coincides with the value reported in Ref.~\cite{gastiasoro2022theory},  $g^\mathrm{TO}_{1,A}=65 \mathrm{meV}\cdot k_F a$. The present results generalize our previous computation for arbitrary polar modes and for all symmetry allowed couplings. 

The coupling constants for other $\bar S_i$ modes can be estimated in an analogous way, by including the appropriate reduced mass for each mode, and the optical gap $\omega_{\bm{q}\alpha}$ of the mode we are interested in. For instance, for a pure $\bar S_2$ mode the only non-zero contribution in Eq.~\eqref{eq:galpha} is $\varphi_{\alpha 2}=\sqrt{\frac{m_u}{\mu_{\bar S_2}}}=0.148$ where $\mu^{-1}_{\bar S_2}=\left(m^\mathrm{Sr}\right)^{-1}+\left(m^\mathrm{Ti}+3m^\mathrm{O}\right)^{-1}$ is the reduced mass of the $\bar S_2$ mode. 

\begin{table}
\caption{Electron-TO coupling of the soft FE mode $\alpha=\mathrm{TO}$ to electronic band $n$ of STO as obtained from Eqs.~\eqref{eq:galphakq}-\eqref{eq:galpha}, assuming a Slater mode Eq.~\eqref{eq:gS1} ($\bar S_1$), a mode from neutron scattering~\cite{Harada1970} Eq.~\eqref{eq:neutron} ($\mathcal{N}$), and from Raman spectroscopy~\cite{Vogt1988} Eq.~\eqref{eq:raman} ($\mathcal{R}$). We have used the experimental phonon frequencies~\cite{Yamanaka2000} at low-$T$ $\hbar\omega_\mathrm{TO}^{E_u}=1$ meV and $\hbar\omega_\mathrm{TO}^{A_{2u}}=2$ meV, and electronic momentum $k_F a=0.3$.}
    \begin{ruledtabular}
    \begin{tabular}{c|ccc|ccc|ccc}
         & & $g^\mathrm{TO}_{n,A}$ [meV]& & & $g^\mathrm{TO}_{n,B}$ [meV] & & & $g^\mathrm{TO}_{n,C}$ [meV] &  \\
       $n$ & $\bar S_1$ & $\mathcal{N}$ & $\mathcal{R}$ & $\bar S_1$ & $\mathcal{N}$ & $\mathcal{R}$ & $\bar S_1$ &$\mathcal{N}$ & $\mathcal{R}$ \\ 
        \hline
       1  & 19 & 21 & 51  & 11 & 10 & 8 & 5 & 5 & 5\\
       2 & -21 & -22 & -47 & -13 &  -16 &  7 & 7 & 9 & 1\\
       3 & 12 & 12 & 23 & 15 & 18 & -5 & -10 & -12 & -4\\
    \end{tabular}
    \end{ruledtabular}
    \label{tab:gcoupling}
\end{table}

Let us now turn instead to a more realistic eigenvector in Eq.~\eqref{eq:expansionphi}, which generally will have a contribution from different $\bar{S}_i$ modes. We can use the normalized atomic displacements estimated for the soft ferroelectric mode by neutron scattering~\cite{Harada1970} and hyper-Raman~\cite{Vogt1988} experiments in the high-$T$ cubic structure. According to these works the $\varphi_{\alpha i}$ coefficients for the expansion in Eq.~\eqref{eq:expansionphi} for the soft ferroelectric mode $\alpha=\mathrm{TO}$ are respectively: 
\begin{align}
\label{eq:neutron}
   & (\varphi_{\alpha 1},\varphi_{\alpha 2},\varphi_{\alpha 3})_\mathrm{neutron}=(0.189,0.059,0.0014),\\
   & (\varphi_{\alpha 1},\varphi_{\alpha 2},\varphi_{\alpha 3})_\mathrm{Raman}=(0.198,-0.0154,-0.08).
   \label{eq:raman}
\end{align}
As already mentioned, both have a predominant $\bar S_1$ mode contribution, but while the first case, Eq.~\eqref{eq:neutron}, implies a small motion of the Sr atoms and a mostly octahedral motion of the oxygens ($|\varphi_1|>|\varphi_2|\gg |\varphi_3|$), the second case, Eq.~\eqref{eq:raman}, suggests a mode in which the Sr atoms are essentially at rest with a significant distortion of the oxygen octahedra ($|\varphi_1|>|\varphi_3|\gg |\varphi_2|$) [see Figs.~\ref{fig:Si-sketch}(b)-(c)].

In order to estimate the Rashba coupling constants arising from these two cases we assume the soft mode in the low-$T$ tetragonal phase is weakly changed and well described by the decomposition with coefficients $\varphi_{\alpha i}$ given by Eqs.~\eqref{eq:neutron}-\eqref{eq:raman}. Substituting these into the weighted sum of Rashba-couplings in Eq.~\eqref{eq:galpha}, we obtain a new set of electron-TO couplings $g^\mathrm{TO}_{n,l}$ for the neutron ($\mathcal{N}$) and Raman ($\mathcal{R}$) eigenvectors. Figure~\ref{fig:lambda}(b) and Table~\ref{tab:gcoupling} collect all the estimated el-TO Rashba-like couplings in this work.
As seen, while the values from the eigenvector from neutron data [Eq.~\eqref{eq:neutron}] and the pure $\bar S_1$ Slater mode are very similar, the resulting couplings are quite different for the eigenvector consistent with Raman data [Eq.~\eqref{eq:raman}]. Indeed, the substantial variation of the el-TO coupling constants in the latter case originates from the intermediate contribution of the oxygen cage distortion of the $\bar S_3$ mode (through the $\varphi_{\alpha 3}$ coefficient) which couples to the gigantic Rashba coefficients $\tau^{(\bar S_3)}_{n,l}$ [see Fig.~\ref{fig:lambda}(a)]. In particular, as shown in Fig.~\ref{fig:lambda}(b),  the absolute value of the $E_u$ coupling $g_{n,A}$ of the Raman determined eigenvector ($\mathcal{R}$) has more than doubled for all three bands and clearly dominates over the other two couplings ($|g_{n,A}|\gg |g_{n,B}|,|g_{n,C}|$), which have been significantly reduced in most cases. We remind the reader that the coupling  $g_{n,A}$ corresponds to the case in which the pseudospin is aligned in the $z$ direction [c.f. Eq.~\eqref{eq:galphakq}].

As a consistency check, we have recomputed the coupling to the Raman mode directly from the band splittings, imposing its eigenvector in a frozen phonon computation in DFT and obtained the same results as with the weighted sum of Rashba-couplings [Eq.~\eqref{eq:galpha}].

Crucial for our results is the weight of the $\bar S_3$ component.  One sees that the oxygen cage appears very rigid in neutrons while it deforms substantially in Raman.
Theoretically, the determination of the soft-mode eigenvector requires the solution of a highly non-harmonic dynamical phonon problem which goes beyond our present scope. As a proxy for this eigenvector, we can examine the fully relaxed broken symmetry ground state, which is polar, since Born-Oppenheimer (adiabatic) DFT does not contain the quantum fluctuations which make the system disorder~\cite{Edge2015}. Such DFT eigenvector has a $\bar S_3$ component similar to that determined by Raman, 
\begin{equation}\label{eq:dfteigen}
    (\varphi_{\alpha 1},\varphi_{\alpha 2},\varphi_{\alpha 3})_\mathrm{DFT}=(0.199,0.014,-0.06)
\end{equation}
This suggests that the Raman determination is more reliable than the one from neutrons, and hence we will consider its eigenvector in the following computations. 

\section{ The superconducting dome}
\label{sec:SC}

%The sensitivity of the couplings to the form of the polar phonon eigenvector has important consequences for superconductivity.
In this Section, starting from our findings on the Rashba electron-phonon interaction in STO presented in Section~\ref{sec:el-ph}, we extend them to high momentum and explore the
consequences for superconductivity. 

One important result from the previous section is that the electron-phonon matrix-elements are very sensitive to the form of the eigenvector of the polar mode. To address this sensitivity we estimate the BCS pairing coupling constant for both the Slater and the Raman determined eigenvectors. For the shake of comparison we restrict now to $E_u$ Rashba couplings of the lowest band ($n=1$),  and the $l=A$ irrep component. This is well justified for the Raman eigenvector since $|g_{n,A}|\gg |g_{n,B}|,|g_{n,C}|$  and partially justified
also for the Slater mode since  $|g^\mathrm{TO}_{1,A}|> |g^\mathrm{TO}_{1,B}|,|g^\mathrm{TO}_{1,C}|$.
%, we will use it to estimate a lower bound of the superconducting coupling constant, which is proportional to the square of the Rashba electron-phonon matrix element~\cite{gastiasoro2022theory}, $\lambda_{\mathrm{BCS},A}=\frac{2}{3} N_F\frac{|g_A^\mathrm{TO}|^2}{\omega_\mathrm{TO}}$. 
Since the magnitude of this matrix element is substantially larger for the Raman determined eigenvector than for the pure $\bar S_1$ mode  [see Fig.~\ref{fig:lambda}(b)] and the superconducting coupling constant is proportional to the square of the Rashba electron-phonon matrix element~\cite{gastiasoro2022theory},
\begin{equation}\label{eq:factor7}
    |g^{\mathrm{TO}(\mathcal{R})}_{1,A}| =2.6|g^{\mathrm{TO}(\bar S_1)}_{1,A}| \longrightarrow \lambda_{\mathrm{BCS},A}^{(\mathcal{R})}\approx 7 \lambda_{\mathrm{BCS},A}^{(\bar S_1)},
\end{equation}
this translates into a factor of 7 larger SC coupling when taking the square.
Hence we see that the details of the eigenvector can strongly influence the resulting bare couplings and in turn its pairing coupling strength. 
For a pure Slater mode, the other $l=B,C$  couplings should also be taken into account and the resulting gap structure and final pairing coupling constant will 
%of course also be affected by changes in 
depend on their ratio~\cite{Sumita2020}. 

Given the DFT determined fully relaxed polar state [Eq.~\eqref{eq:dfteigen}], 
in the following we will assume the soft-mode is best described by the hyper-Raman determined eigenvector Eq.~\eqref{eq:raman}. Luckily, this simplifies computations since $|g_{n,B}|$ and $|g_{n,C}|$ can be neglected,
and estimates based on the sole contribution of the $l=A$ irrep component are well justified. 

We also note that we will focus solely on the $s$-wave superconducting channel. It has been found by different groups~\cite{Kozii2015,kozii2019,Gastiasoro2020,Sumita2020} that the odd-in-$k$ Rashba mechanism has attractive higher angular momentum Cooper channels ($p$-wave, $d$-wave etc.), but sub-leading to the $s$-wave channel in cubic and tetragonal systems. We therefore restrict our high momentum study to $s$-wave pairing solutions.

\subsection{Generalized Rashba in DFT and dome behavior}

So far we have explored the conventional Rashba-like $k$-linear model in Eq.~\eqref{eq:Gamman}, which describes the coupling between the soft FE phonon and the electrons fairly well at low momenta, as we have shown by frozen phonon computations [Fig.~\ref{fig:splitdft}]. However, as already mentioned in Section~\ref{sec:lineardft}, the band split obtained by the \emph{ab initio} computations exhibits deviations from linear-in-$k$ beyond a characteristic momenta $k a$ which generally depends on the electronic band $n$, the momentum direction $\hat{\bm{k}}$ and the polar mode $\bar S_i$. 

\begin{figure}
    \centering
    \includegraphics[width=\linewidth]{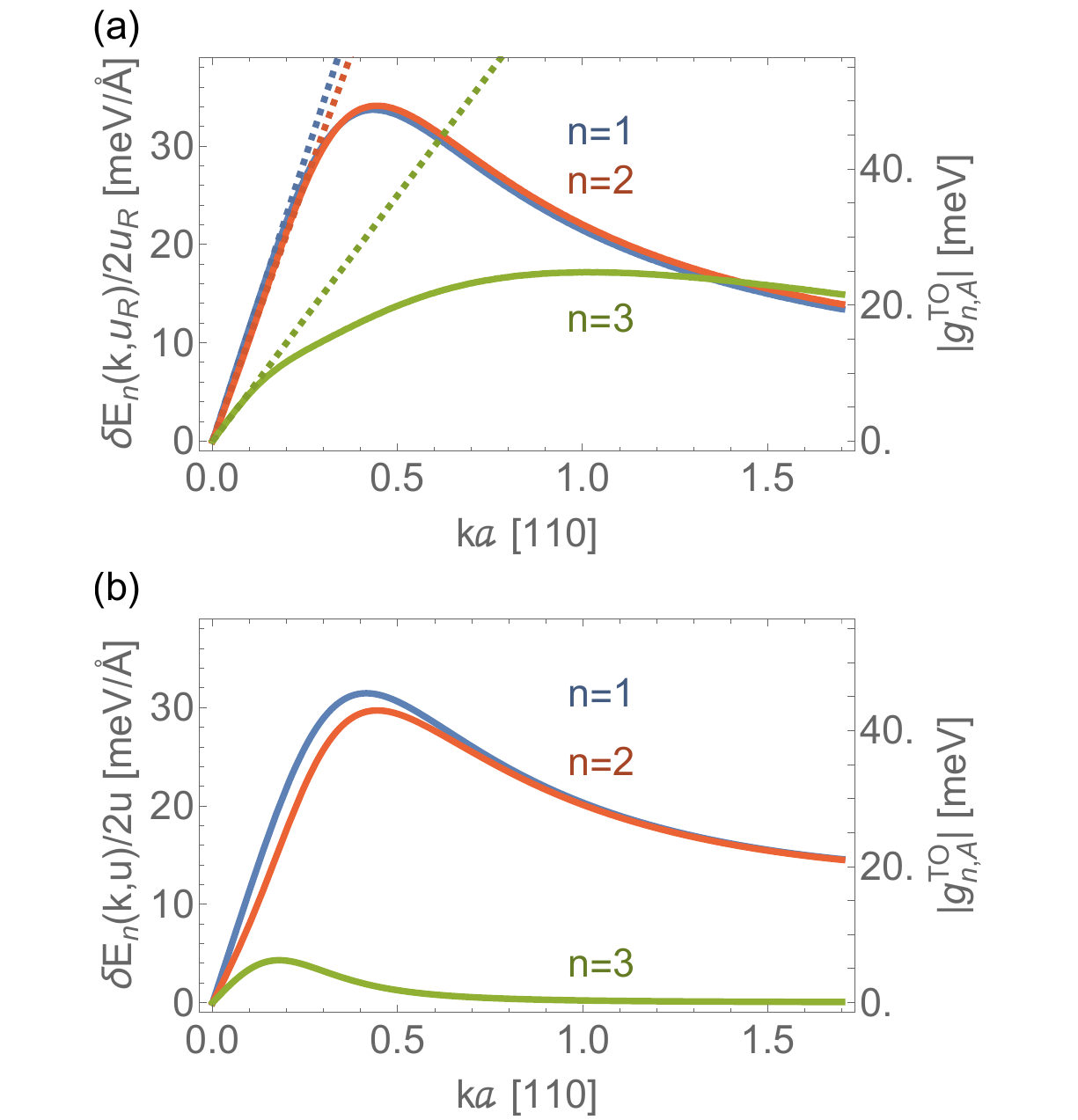}
    \caption{ Electronic band split for the three bands of STO in the presence of a frozen polar mode with eigenvector deduced from Raman [Eq.~\eqref{eq:raman}] and with polar axis $\bm{\hat{n}}_p^{E_u}\parallel[1\bar{1}0]$
  and momentum $\bm{k}\parallel[110]$.  The right $y$-axis shows the corresponding generalized Rashba electron-phonon matrix element [Eq.~\eqref{eq:galphabeyond}].
Solid lines in (a)  are \emph{ab initio} results and dashed lines the conventional Rashba model [Eq.~\eqref{eq:deltaEnuy} and Eq.~\eqref{eq:galpha}].
Panel (b) is the result for minimal $t_{2g}$ model in Eqs.~\eqref{eq:Hband}, \eqref{eq:Hu} and \eqref{eq:txy0}  with the coupling matrix element corresponding to the Raman determined eigenvector, $2\partial_x t_{xy0}=\tau^\mathcal{R}_{1,A}=115$ meV/{\AA} (extracted from DFT results in (a)).}
    \label{fig:kfRaman}
\end{figure}

Figure~\ref{fig:kfRaman}(a) shows the \emph{ab initio} results (solid lines) of the pseudospin-split of each $n$ band going beyond the small $k a$ values presented in Fig.~\ref{fig:splitdft}. We chose a frozen-phonon $E_u$ Raman mode [Eq.~\eqref{eq:raman}] with polar axis $\hat{\bm{n}}_p\parallel[1\bar 1 0]$, and show the band split along the perpendicular momentum direction $\hat{\bm{k}}\parallel[110]$. 
As seen, for all bands the band-split $\delta E_n$ is initially conventional Rashba-like (growing linearly with momenta), but deviates from linearity and peaks at intermediate values of momenta after which steadily decreases in a form close to $1/k$. 

To linear-in-$k$ order the splittings are given by the Rashba couplings through Eq.~\eqref{eq:deltaEnuy} and Eq.~\eqref{eq:galpha}. 
We can generalize Eq.~\eqref{eq:galpha} to an arbitrary odd function of $k$ by introducing  for each electronic band $n$,
\begin{equation}
\label{eq:galphabeyond}
    g^{\bm{q}\alpha}_{n,l}= k a \mathcal{F}_{n,l}( k a) \sqrt{\frac{\hbar}{2m_{u}\omega_{\bm{q}\alpha}}}\tau^{(\alpha)}_{n,l},
\end{equation}
were we defined  $\mathcal{F}_{n,l}( k a)$ such that it is even in $k$ and $\mathcal{F}_{n,l}( k a)\rightarrow 1 $ for $k\rightarrow0$.  
By definition, the electron-phonon matrix element is proportional to the band split so $\mathcal{F}_{n,l}(k a)$ was extracted directly from the \emph{ab initio} results. The corresponding $g_{n,A}^\mathrm{TO}$ is given by the right $y$-axis in Fig.~\ref{fig:kfRaman}(a). We anticipate that this dome in $k$  results in a dome in electronic density for both $\lambda_{\mathrm{BCS}}$ and $T_c$ so it is important to discuss its origin, which we do next.

\subsection{Minimal model and dome behavior}\label{sec:minmoddome}
The dome-like behavior of the band split can be traced back to a $k$-dependent quenching of angular momentum. The   essential physics is  captured by the minimal model in Eq.~\eqref{eq:tb-model} supplemented by a 
one-parameter simplification of the polar interaction in Eq.~\eqref{eq:Hu}. Namely we keep only the spin-conserving $j=0$ term and restrict to mixing of $x$ and $y$ orbitals, 
\begin{equation}
\label{eq:txy0}
  t_{xy0}(\bm{k},\bm{q})= 2i\partial_x t_{xy0} u(\bm{q}) \left[-\sin\left(k_ya\right)\hat{n}_{px}+\sin\left(k_xa\right)\hat{n}_{py}\right] 
\end{equation}
Computing the band splitting along $\hat{\bm{k}}\parallel[110]$ for 
$\hat{\bm{n}}_p\parallel[1\bar 1 0]$ one obtains panel (b) of Fig.~\ref{fig:kfRaman}. As seen, this simple approximation captures very well the coupling of the first two bands including the dome behavior. It underestimates the coupling of the third band which therefore calls for additional parameters beyond the scope of this subsection.  

\begin{figure}
    \centering
    \includegraphics[width=\linewidth]{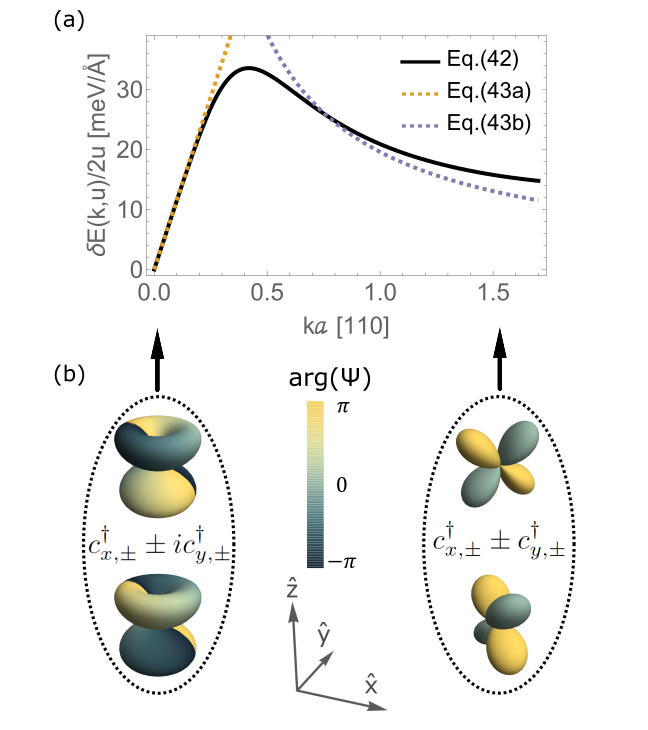}
    \caption{(a) Band split for two-orbital toy-model Eq.~\eqref{eq:deltaEanal}, and the perturbative limits Eqs.~\eqref{eq:deltaEanalpert1}-\eqref{eq:deltaEanalpert2}.
     (b) $k$-dependence of eigenstates, in the perturbative limits shown in (a). The color legend represents the argument of the wave function.}
    \label{fig:toymodel}
\end{figure}

Focusing on the lowest band, 
the wave function near $\Gamma$ [Eq.~\eqref{eq:c3232}], suggests to simplify even more the tight-binding $t_{2g}$ model Eq.~\eqref{eq:tb-model} by restricting it to two orbitals 
$\mu=x,y$, which is formally equivalent to taking the AFD parameter to infinity, $\Delta\rightarrow\infty$ in Eq.~\eqref{eq:adf}. 
%Note also that for an in-plane polar displacement $\hat{\bm{n}}_p\parallel[1\bar 1 0]$ and a band split along $\hat{\bm{k}}\parallel[110]$ [Fig.~\ref{fig:kfRaman}(a)], only the spin-conserving inter-orbital hopping elements  are left in Eq.~\eqref{eq:tauxy}. That is, the relevant symmetry breaking term in the polar coupling Hamiltonian Eq.~\eqref{eq:Hu} to linear order in the polar amplitude $u$ is
The band split $\delta E$ of this toy-model for a finite $u(\bm{q})$ and the same $\hat{\bm{k}}$ and $\hat{\bm{n}}_p$ orientations as above can be analytically obtained from the eigenvalues of a $2\times 2$ matrix with matrix elements,
\begin{align*}
    h_{11}&=h_{22}=2\left(t_1+t_2+2 t_3\right)\left[1-\cos\left(\frac{ka}{\sqrt{2}}\right)\right] \\
    h_{12}&=h_{21}^*=i\xi -4t_4\sin^2\left(\frac{ka}{\sqrt{2}}\right) +2i \partial_x t_{xy0}u\sqrt{2}\sin\left(\frac{ka}{\sqrt{2}}\right)
\end{align*}
and reads 
\begin{equation}
\label{eq:deltaEanal}
    \frac{\delta E(k[110],u[1\bar{1}0])}{2u}=\frac{ \tau_{1,A}\sqrt{2}\sin\left(\frac{ka}{\sqrt{2}}\right)}{\sqrt{1+\left[\frac{4t_4\sin^2\left(\frac{ka}{\sqrt{2}}\right)}{\xi}\right]^2}},
\end{equation}
where we have used the relation $2\partial_x t_{xy0}=\tau_{1,A}$  [ Eq.~\eqref{eq:lambdan1}].
Expanding to linear order in $k$ one recovers the conventional Rashba form of Ref.~\cite{gastiasoro2022theory} 
%(where we used the notation $t'_{xy}$ for $\partial_x t_{xy0}$)
which here is generalized to arbitrary momentum.

Equation~\eqref{eq:deltaEanal} is plotted in Fig.~\ref{fig:toymodel}(a) for the electronic parameters in STO $t_4$ and $\xi$ listed in Sec.~\ref{sec:minmod}, and $\tau^{\mathcal{R}}_{1,A}=115$ meV/{\AA} corresponding to the Raman deduced soft-mode eigenvector. As seen, the two-orbital toy model excellently captures the band split of the lowest band, $n=1$, computed by DFT [c.f. Fig.~\ref{fig:kfRaman}(a) and Fig.~\ref{fig:toymodel}(a)]. Furthermore, the analytical result for $n=2$ is identical to the $n=1$ case while in DFT both are very similar. Thus, surprisingly, the two-orbital toy-model provides a good approximation to the second band despite its non-negligible weight of the $z$-orbital near $\Gamma$ [Eq.~\eqref{eq:c3212}]. This is attributed to the rapid decrease of the $z$-orbital character as momentum increases along $[110]$. Indeed, including the $z$ orbital results in little change on the splittings for $n=1,2$ [Fig.~\ref{fig:kfRaman}(b)].

The $k$-dependence in Eq.~\eqref{eq:deltaEanal} is determined by the competition between the SOC energy $\xi$ in $\mathcal{H}_\mathrm{SOC}$ (which dominates at $k=0$), and the hopping term $4t_4\sin^2\left(\frac{ka}{\sqrt{2}}\right)$ [Eq.~\eqref{eq:tmunu}] which induces a mass mismatch of the bands in $\mathcal{H}_0$, and increases with $k$. 
The former term promotes a state with $l_z=\pm1$ angular momentum $c^\dagger_{x,\pm}\pm i c^\dagger_{y,\pm}$ [Eq.~\eqref{eq:c3232}] leading to 
$j_z=\pm3/2$ states, whereas the latter term constrains the system towards $c^\dagger_{x,\pm}\pm c^\dagger_{y,\pm}$ states which have $\langle {\bm l}\rangle=0$. This $k$-dependent quenching of angular momentum is illustrated in Fig.~\ref{fig:toymodel}(b) where the complex, $l_z=\pm1$ (real,  $\langle \bm{l}\rangle=0$) orbitals for small (large) $k$ are shown.   
%Note that the specific form of the hopping energy term competing with SOC (which in the case we illustrated is $4t_4\sin^2\left(\frac{ka}{\sqrt{2}}\right)$) will generally depend on momentum direction $\hat{\bm{k}}$, and the band involved.

Doing perturbation in Eq.\eqref{eq:deltaEanal} in the two opposite limits, where the SOC term dominates over the hopping term and vice versa, one obtains the following expressions for the band split in the continuum limit ($ka\ll 1$), %Eq.~\eqref{eq:deltaEanal},
%\begin{subnumcases}{\frac{\delta E}{2u}\approx} 
%   \tau_{1,A} \sqrt{2} \sin\left(\frac{ka}{\sqrt{2}}\right)   +\mathcal{O}\left(\frac{4t_4\sin^2\left(\frac{ka}{\sqrt{2}}\right)}{\xi}\right)^2 \\   \frac{\sqrt{2} \tau_{1,A}\xi}{4t_4\sin\left(\frac{ka}{\sqrt{2}}\right)} +\mathcal{O}\left(\frac{\xi}{4t_4\sin^2\left(\frac{ka}{\sqrt{2}}\right)}\right)^2
%\end{subnumcases}  

   %+\mathcal{O}\left(\frac{t_4^2 (ka)^3 }{\xi^2}\right) \\
   
\begin{subnumcases}{\frac{\delta E}{2u}\approx}
     \tau_{1,A} ka  \left[1+\mathcal{O}\left(\frac{t_4^2 (ka)^4 }{\xi^2}\right) \right],\ \ \ \ t_4(ka)^2\ll\xi \label{eq:deltaEanalpert1}\\
     \frac{\tau_{1,A}\xi}{2t_4ka}\left[1+
%     \frac{(ka)^2}{12}+
     \mathcal{O}\left(\frac{\xi^2}{t_4^2 (ka)^4}\right)\right],\ \ \ \ \xi\ll t_4(ka)^2
    \label{eq:deltaEanalpert2}
\end{subnumcases}
We recover the Rashba linear-in-$k$ term Eq.~\eqref{eq:deltaEnuy} when the SOC energy term dominates over the hopping term [Eq.~\eqref{eq:deltaEanalpert1}], and the $1/k$ dependence in the opposite limit, when the kinetic term takes over [Eq.~\eqref{eq:deltaEanalpert2}]. These two perturbative expressions are shown in Fig.~\ref{fig:toymodel}(a) together with the full expression Eq.~\eqref{eq:deltaEanal}, by dashed orange and purple lines, respectively.
Deviations of the expansion at large momentum are due to lattice effects which were neglected as they do not change the qualitative picture. 
Clearly, when the angular momentum becomes quenched the spin-orbit assisted electron-phonon interaction becomes ineffective and dies out. 
In the continuum limit the maximum of the coupling is given by
\begin{equation}
     k^{max} a=\sqrt{\frac\xi{2t_4}}\approx 0.42,
\end{equation}
which again illustrates the competition between spin orbit and band mass mismatch energies. The right hand side corresponds to the present parameters, relevant for STO.  As it will be clear below, the maximum of the coupling is the more important factor to determine the optimum Fermi momentum and density for superconductivity. 

\begin{figure}
    \centering
    \includegraphics[width=\linewidth]{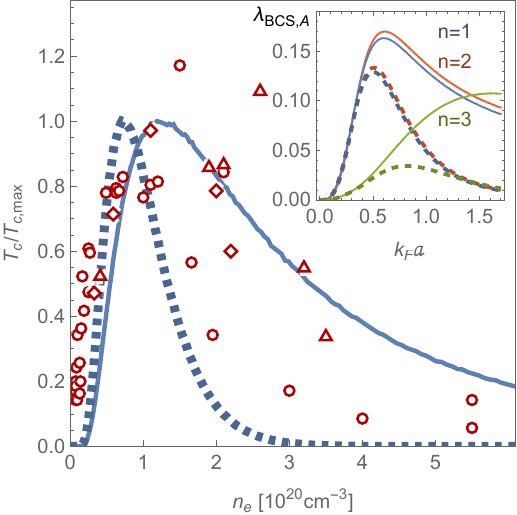}
    \caption{$T_c$ dome (normalized to its maximum value) vs. carrier density $n_e=\frac{k_F^3}{3\pi^2}$  assuming a parabolic band and using $\lambda_{\mathrm{BCS},A}$ from lowest band $n=1$ shown in inset. Full (dashed) line neglects (includes) the hardening of the TO mode. Inset: band resolved $\lambda_{\mathrm{BCS},A}= \frac{2}{3} N_F \frac{|g_{n,A}^\mathrm{TO}|^2}{\omega_\mathrm{TO}}$ using the \emph{ab initio} results $g_{n,A}^\mathrm{TO}$ of Fig.~\ref{fig:kfRaman}(a). 
    Open symbols are bulk $T_c$ experimental data from Ref.~\cite{Koonce1967} (circles), Ref.~\cite{Collignon2017}(triangles) and Ref.~\cite{thiemann2018single} (diamonds) with $T_{c,max}=0.35$K. Notice that for simplicity an effective one-band is considered in the theory to compute $n_e$, while in the experiment $n_e$ is the total electronic density.
}
    \label{fig:tc}
\end{figure}

Because the pairing interaction arising from the polar coupling is in turn proportional to the square of the electron-phonon matrix element~\cite{gastiasoro2022theory} (shown in the right $y$-axis of Fig.~\ref{fig:kfRaman}), $V_\mathrm{TO}\propto|g^\mathrm{TO}|^2$, it also acquires a pronounced peak as a function of $k=k_F$, the Fermi momentum. The initial quadratic increase with $k_F$ peaks and decreases as $V_\mathrm{TO}\propto 1/k_F^2$ for all three bands. 
Consequently the pairing coupling constant\cite{gastiasoro2022theory} $\lambda_{\mathrm{BCS},A}=\frac{2}{3}N_F V_\mathrm{TO}\propto k_F V_\mathrm{TO}$ of each band (assuming parabolic bands) shows a dome-like form with increasing $k_F$ [inset of Fig.~\ref{fig:tc}]. We took a constant factor of 2 effective mass enhancement in $N_F$, chosen to match the values from specific heat~\cite{McCalla2019} at low carrier densities.
This mass renormalization can be viewed as effectively taking into account the coupling to other phonons not considered explicitly so far, such as the longitudinal optical modes. 
%In any case, our choice of a fixed factor of 2 mass enhancement for all fillings can be considered a lower bound for $\lambda_{\mathrm{BCS},A}$.

The obtained values of $\lambda_{\mathrm{BCS},A}$ are in the weak coupling limit so we can use a simple BCS formalism.  This is in agreement with the $2\Delta(T=0)/T_c$ ratio obtained with tunneling and microwave spectroscopy studies~\cite{thiemann2018single,swartz2018polaronic} which suggest that a weak coupling picture applies.
Neglecting inter-band (finite-$q$) couplings in Eq.~\eqref{eq:Husoc}, 
%(as we are studying the $\bm{q}\rightarrow \bm{0}$ polar phonon contribution to pairing at the Fermi surface), 
we obtain three uncoupled SC gap equations, one for each electronic band. In this approximation, the higher $T_c$ corresponds to the bulk critical temperature of the material. 

This simplified picture already predicts a dome of $T_c\propto \exp [-1/\lambda_{BCS,A}]$ within the generalized Rashba coupling pairing mechanism, shown in Fig.~\ref{fig:tc} for the first band $n=1$. Notice that the largest $T_c$ corresponds to $n=2$, but since both curves are very similar there is little difference on which one is chosen.
Note also that we are assuming a rigid band picture, which seems to be a good approximation for bulk SC obtained by Nb and La doping~\cite{fauque2022electronic}.

Within the present mechanism, we believe that the dominant correction to the above computation of the dome is given by the hardening of the TO mode with carrier density~\cite{Gastiasoro2020review,yu2021theory}, $\tilde\omega_\mathrm{TO}(k_F a)=\sqrt{\omega_\mathrm{TO}^2+D ( k_F a)^3}$ with $D=1.64 \mathrm{ meV}^2$. This parametrization was chosen to match the factor of three hardening of the soft mode at low temperatures when reaching $n_e=1.5\times 10^{20}$ as measured by infra-red spectroscopy\cite{Devreese2010}.  
As seen by the dashed lines in Fig.~\ref{fig:tc}, since $\lambda_\mathrm{BCS}\propto 1/\omega^2_\mathrm{TO}$, the hardening reduces the pairing constant even faster at high Fermi momenta where the hardening effect is largest. Consequently, the $T_c$ dome is narrowed and shifted to lower densities.

\subsection{Comparison with experiments}

A key prediction of the present mechanism is the density where $T_c$ peaks, which including (neglecting) the hardening of the soft mode is found to be $n^{opt}_e\approx 7\times 10^{19}$~cm$^{-3}$ ($n^{opt}_e\approx 1\times 10^{20}$~cm$^{-3}$ ). This can slightly change by multiband effects and mass anisotropy but the order of magnitude  is in excellent agreement with 
%the optimum $T_c$ from 
diamagnetic experiments\cite{Schooley1965,Koonce1967,thiemann2018single},  without free parameters. To the best of our knowledge, the fine resolution in carrier density achieved in early Refs.~\cite{Schooley1965,Koonce1967} has not since been attained in bulk superconductivity measurements. To determine the experimental optimum density, we have neglected two outlier points in Ref.~\cite{Koonce1967} 
that were neglected also in their fit and a similar outlier point in the data of Ref.~\cite{Collignon2017}, which unfortunately does not cover the maxima of Refs.~\cite{Schooley1965,Koonce1967,thiemann2018single}.
More experimental and theoretical work is needed to understand if those outliers are a systematic effect. Eventually, they may be attributed to multiband effects neglected here. Furthermore, we are not taking into consideration resistivity data, as 
it is not a bulk superconductivity probe. Indeed, 
$T_c$ in bulk probes (specific heat, thermal conductivity and diamagnetism) is consistently lower than the zero-resistance $T$, which points towards filamentary superconductivity at higher temperatures and extremely dilute samples~\cite{collignon2019,bretz2019superconductivity,fauque2022electronic}. 

When plotted in linear scale, it becomes clear that the dome is very asymmetric, with a rapid rise and a much slower decrease [Fig.~\ref{fig:tc}]. Also, this asymmetry is well reproduced by the theory presented here. The rise of $T_c$ appears at a density slightly higher than in experiments while the decrease is very sensitive to what is assumed for the hardening of the soft mode. Within the present uncertainties on the experimental data [Fig.~\ref{fig:tc}], also the width and the asymmetry of the dome are in very good agreement with experiment.

An important remaining question is if the present theory can explain the observed maximum value of $T_c$. 
In Ref.~\cite{gastiasoro2022theory} we used the naive $k$-linear Rashba form, neglected the phonon hardening, and obtained a good 
estimate of $T_c$ near optimum density.  The present computations show that both approximations can lead to an overestimation of $\lambda_\mathrm{BCS}$. On the other hand, a pure Slater mode $\bar{S}_1$ was assumed; Eq.~\eqref{eq:factor7} shows that the details of the soft mode eigenvector strongly affects the coupling strength. For the soft mode eigenvector found by hyper-Raman   [Eq.~\eqref{eq:raman}], the overestimation error of Ref.~\cite{gastiasoro2022theory} tends to cancel with the amplification of Eq.~\eqref{eq:factor7}
so the computed $\lambda_\mathrm{BCS}$ is again close to the one required by experiments. Indeed, taking the BCS form $k_B T_c=1.13 \hbar \omega_{TO} \exp[-1/\lambda_\mathrm{BCS}]$ and assuming the soft mode frequency to be $\omega_{TO}=3$ meV near the optimum density requires a $\lambda_\mathrm{BCS}=0.21$ to obtain a maximum transition temperature near the experimental one\cite{Schooley1965,Koonce1967,thiemann2018single,collignon2019} ($T_{c,max}\approx 0.35$ K).
In our computation the maximum $\lambda_\mathrm{BCS}$, including phonon hardening, is $\lambda_{max}=0.13$  [see inset of Fig.~\ref{fig:tc}] which is fairly close to this BCS estimate without any free parameters. Thus, it is clear that the present mechanism can explain the observed $T_c$ as $\lambda_{max}$ is in the correct range.  

As explained above, our estimate of $\lambda_\mathrm{BCS}$ is a lower bound. A more accurate computation should take into account that around optimum carrier density, three bands are filled; our estimate takes into account only one band. Also, the contributions of $g_{n,B}$ and $g_{n,C}$ matrix elements to the $A_{1g}$ pairing channel have been neglected. Both of these aspects are expected to increase $T_c$ and hence improve the agreement with the experiment. 
While one can incorporate the above corrections in the computations, in practice, due to the exponential dependence of $T_c$ with $\lambda_\mathrm{BCS}$ and the approximate treatment of the Coulomb interaction~\cite{marsiglio2022impact} (neglected in this work), a high level of accuracy in $T_c$ should not to be expected. 
%practically never achieved in  materials which are in the weak coupling regime, not even for well understood materials. 
Therefore, for the time being, we consider the present results robust enough to claim that the magnitude of $T_c$ can be explained within the present mechanism.
As mentioned earlier, the coupling of electrons to pairs of TO modes, that is, the quadratic coupling to the FE mode, is also a promising source of pairing in doped paraelectrics, and certainly in doped STO~\cite{Ngai,van2019possible,kiseliov2021theory,volkov2021superconductivity,zyuzin2022anisotropic}. Also, for this mechanism, a dome-like feature of $T_c$ vs. doping 
and an optimum density in good agreement with experiment has been found without free parameters~\cite{volkov2021superconductivity}. Therefore, more experimental and theoretical work is needed to decide which of the two mechanisms is more appropriate to describe the superconducting dome. 

We remark that the dome arising in Ref.~\cite{volkov2021superconductivity} is due to the hardening of the phonons with electronic carrier density, with the quadratic coupling constant in the interaction vertex left as a constant. In the work presented here, on the other hand, the momentum dependence (odd-parity) of the linear coupling constant already gives rise to a dome (see full lines in Fig.~\ref{fig:tc}), and the hardening of the soft phonon modifies its shape (dashed lines  in Fig.~\ref{fig:tc}). 

Charge transport in the normal state of doped STO presents a pronounced $T^2$ regime in the resistivity present even at very low $n_e$ where umklapp scattering is forbidden. This behavior has been explained invoking the combination of scattering by a LO mode and two-TO modes~\cite{kumar2021} as well as a LO mode and a single-TO mode over a range of temperatures~\cite{yu2021theory}. This suggests both linear and quadratic coupling mechanisms are consistent with the observed $T^2$ resistivity. Whether the momentum structure of the linear coupling found in this work can discriminate between the two scenarios for charge transport is left for a future study.

\section{Summary and Conclusions}
\label{sec:conclusions}

In this work, we have derived the most general Rashba-like linear coupling between the electronic bands and the polar modes at the zone center of tetragonal STO [Eq.~\eqref{eq:Gamman}]. Fitting the electronic band split of the Rashba model [Eqs.~\eqref{eq:deltaEnuy}-\eqref{eq:deltaEnuz}] to \emph{ab initio} $q=0$ frozen-phonon calculations [Fig.~\ref{fig:splitdft}] we have estimated the corresponding Rashba couplings, $\tau^{(\bar S_i)}_{n,l}$, to zone center polar modes $\bar S_i$ [Fig.~\ref{fig:lambda}(a)]. These $\bar S_i$ modes form a complete basis of in-plane, $E_u$, and out-of-plane, $A_{2u}$, modes at the zone center in STO [Eqs.~\eqref{eq:S1}-\eqref{eq:S5}] and hence we have mapped out the entire Rashba-like linear coupling subspace of these zone-center polar modes.  

The origin of the Rashba couplings can be understood as arising from induced hopping channels between neighboring $t_{2g}$ orbitals in the presence of polar distortions [Eqs.~\eqref{eq:Hu}]. We have explicitly shown how to connect the symmetry allowed Rashba-coupling constants $\tau_{n,l}$ in Eq.~\eqref{eq:Gamman} to the microscopic hopping processes $\partial_l t_{\mu\nu}$ [Eqs.~\eqref{eq:tauxy} and \eqref{eq:lambdan1}].  Indeed, a minimal, three-orbital model with only one induced hopping parameter reproduces qualitatively and to some extent quantitatively many results of the {\it ab initio} computations [Fig.~\eqref{fig:kfRaman}]. 

We have shown how to estimate the electron-polar-phonon coupling function of a general polar mode by decomposing it into the $\bar S_i$ basis [Eqs.~\eqref{eq:galphakq}-\eqref{eq:expansionphi}] using the obtained Rashba-couplings $\tau^{(\bar S_i)}_{n,l}$ [Fig.~\ref{fig:lambda}(a)]. Since an accurate determination of the soft-mode eigenvector is difficult, this allows us to separate the problem of determining the coupling to the soft mode from the problem of determining its eigenvector.

Starting from eigenvector cases of a Slater mode $\bar S_1$ and those consistent with neutron data [Eq.~\eqref{eq:neutron}] and hyper-Raman data [Eq.~\eqref{eq:raman}] in STO, we have obtained and compared three sets of electron-TO-phonon couplings $g_{n,l}^\mathrm{TO}$ [Fig.~\ref{fig:lambda}(b) and Table~\ref{tab:gcoupling}]. We find substantial coupling values for the three electronic bands, with the $E_u$ coupling $g_{n,A}^\mathrm{TO}$ being larger than the other two couplings $g_{n,B}^\mathrm{TO}$ and $g_{n,C}^\mathrm{TO}$. 
Physically, the $A$ component of the $E_u$ irrep corresponds to the case in which the pseudospin lies  along the $z$-direction, as considered in Ref.~\cite{gastiasoro2022theory} and in the toy-model of Sec.~\ref{sec:minmoddome}.

We found that the details of the eigenvector can substantially alter the magnitude of the couplings. In particular, the intermediate contribution of the $\bar S_3$ mode (which distorts the oxygen octahedra) to the Raman  determined eigenvector [Eq.~\eqref{eq:raman}]  significantly increases the $E_u$ coupling $g_{n,A}^\mathrm{TO}$ for all three bands [see Fig.~\ref{fig:lambda}(b)].
%More measurements in this direction would allow to 
%have a better understanding of the eigenvector of the FE soft mode in STO at low temperatures to 
%narrow down even more the electron-TO-phonon coupling constant values %$g_{n,l}^\mathrm{TO}$. 

Certainly, the sensitivity of the couplings to the eigenvector has important implications for the Rashba pairing mechanism~\cite{gastiasoro2022theory}, since the BCS coupling constant $\lambda_{BCS}$ is proportional to the square of the electron-TO-phonon coupling in its simplest form.
Comparison with the DFT broken symmetry state suggests that the oxygen cage should deform substantially in the soft mode as found by the  hyper-Raman eigenvector. 
More experimental and theoretical work will be highly desirable to refine the eigenvector of the soft mode and improve the estimate of $\lambda_\mathrm{BCS}$. 

While the linear-in-$k$ conventional Rashba model works well at low momenta, our \emph{ab initio} frozen phonon results generally indicate a deviation from the linear-$k$ Rashba split of the three electronic bands beyond a critical wave vector for all polar modes. As a result, we find a dome-like behavior of the electron-TO-phonon coupling $g_{n,l}^\mathrm{TO}$ [Eq.~\eqref{eq:galphabeyond}]: beyond the critical wave vector the linear-$k$ growth slows and peaks, subsequently evolving into a $1/k$ decrease [Fig.~\ref{fig:kfRaman}]. This behavior is due to a $k$-dependent quenching of the angular momentum, as explicitly shown by reducing the three-orbital model to a two-orbital toy model [Eq.~\eqref{eq:deltaEanal} and Fig.~\ref{fig:toymodel}].

Assuming a rigid band shift for electronic doping, and without introducing electronic screening effects, 
this deviation from Rashba already entails a dome for $T_c$ as a function of electronic density. We remark that this result does not depend crucially on the particular form of the polar mode eigenvector.
A popular explanation of the dome invokes the proximity to a ferroelectric quantum critical point~\cite{Edge2015}. In the simplest picture, the dome is attributed to the change  of the frequency of the soft-mode with density, possibly with the structural transition laying below the dome~\cite{setty2022superconducting}.
This last simplified picture, however, requires that the system breaks inversion symmetry to the right or to the left of the optimum density which, to the best of our knowledge, is not the case. 
In the mechanism we presented, the proximity to the ferroelectric quantum critical point is important to have a soft-mode in the first place, which increases $\lambda_\mathrm{BCS}\propto 1/\omega_\mathrm{TO}^2$, but it is not 
 responsible for the non-monotonous behavior of $T_c$. Including the hardening with doping increases the agreement with the experimental data of Refs.~\cite{Koonce1967,Schooley1965} [Fig.~\ref{fig:tc}].

The obtained maximum value of $T_c$, the density at which it peaks, and the characteristic asymmetry of the dome are in surprisingly good agreement with experiments without free parameters [Fig.~\ref{fig:tc}]
providing a compelling solution to a more than 50 year old open problem in the field.  
Small deviations remain, which we attribute to several simplifications we have made to avoid introducing more parameters in the theory. For example, additional Rashba coupling contributions to the 
$A_{1g}$ pairing channel can be considered, which will increase $T_c$ and possibly decrease the smaller density  at which bulk superconductivity becomes robust. It remains an interesting question for future research to find out if the spin-orbit processes neglected [labeled $B$ and $C$ in Eq.~\eqref{eq:Gamman}] can stabilize different pairing symmetries other than the conventional $s$-wave we have considered.

The approach presented here can be generalized to study the linear coupling characteristics between electrons and FE soft TO modes, as well as the corresponding pairing mechanism in other incipient ferroelectrics. A noteworthy example are KTO interfaces, where the linear coupling to the TO mode has recently been invoked for pairing~\cite{liu2022tunable}. 

\section*{Acknowledgements}
We thank P. Volkov for useful discussions.
We acknowledge financial support from the Italian MIUR through Projects No. PRIN 2017Z8TS5B, and No. 20207ZXT4Z. 
M.N.G. is supported by the Marie Skłodowska-Curie individual fellowship Grant Agreement SILVERPATH No. 893943. 
We acknowledge the CINECA award under the ISCRA initiative Grants No. HP10C72OM1 and No. HP10BV0TBS, for the availability of high-performance computing resources and support.

%\appendix

\bibliography{biblio}
% Add references to biblio.bib file

\end{document}